\documentclass[12pt,preprint]{aastex}

\newcommand{\lradio}{\ensuremath{\mathrm{erg~s^{-1}Hz^{-1}}}}
\newcommand{\lf}{\ensuremath{L_{5100}}}

\newcommand{\fwhb}{\ensuremath{\mathrm{FWHM}_\mathrm{H{\beta}}}}

\newcommand{\msun}{\ensuremath{M_{\odot}}}
\newcommand{\mhole}{\ensuremath{M_\bullet}}
\newcommand{\ovr}{\ensuremath{\langle R \rangle}}

\slugcomment{}

\shorttitle{From quasars to radio galaxies}
\shortauthors{Reviglio, Helfand}

\begin{document}

\title{Active Galaxies in the Sloan Digital Sky Survey III: from quasars to radio galaxies?}

\author{Pietro M. Reviglio,  David J. Helfand}
\affil{Astronomy Department, Columbia University, New York, NY 10027}

\email{reviglio@astro.columbia.edu,djh@astro.columbia.edu}

\begin{abstract}

In this third of a series of papers concerning active galaxies in the FIRST and Sloan Digital Sky Surveys, we analyze the spectroscopic and radio properties of a sample of narrow-line  Active Galactic Nuclei (AGN), broad-line Seyfert I galaxies, and Quasars in the local universe in order to investigate the dependence of their activity on the mass, spin and accretion rates  of the supermassive black holes (SMBH) residing at the centers of their host galaxies.

We show that galaxies hosting more massive SMBH are more likely to power stronger and larger radio jets,  and we show a strong anti-correlation between the strength of the lines of radio emitting galaxies and their radio power. 
Furthermore we show that the compactness of a jet is correlated with the epoch of the last episode of star-formation, suggesting a link between the presence of cold gas in a galaxy, the size of its SMBH  and the radio and spectroscopic features of its AGN. 

We use our large statistical sample to test the  expectations  of unified models of AGN based on orientation. While confirming that Seyfert II galaxies and radio galaxies are significantly more extincted then Seyfert Is and nearby Quasars, we find several major inconsistencies with such a paradigm. In particular we show a  strong difference in the  [OIII],[OII] and [NII] luminosities  for different spectroscopic classes, a result  which argues in favor of an evolution of the broad and narrow line regions of active nuclei over time. 

We suggests that  evolution, rather than orientation, may be the key element in shaping the properties of active nuclei, as also suggested by the results of high-redshift X-ray and radio surveys and we speculate on a model that may predict this kind of evolution.

\end{abstract}


\section{Introduction}

Radio emission in active galaxies may hold the key to understanding the formation, growth  and  evolution of active galactic nuclei and the engines widely believed  to power them: supermassive black holes.

In his early paper, Lynden Bell (1969) argued that  low-power active galactic nuclei were simply ``old collapsed quasars'' -- the remnants of what was left after the powerful activity waned and the supermassive black hole was left in a state of lower activity. However, the problems of early evolutionary scenarios \cite[]{Ryle1967, Lynden1969, Rees1982}  in explaining the transformation of nuclear features, in particular the differences in emission lines  in quasars and radio galaxies\cite[]{Hes1993}, have favored over time scenarios where the differences between high- and low-power AGN are interpreted in terms of the relative orientation of the nuclear engine and the observer \cite[]{Antonucci1993, Urry1995}.

However, evolution in radio AGN has been known for many years \cite[]{Longair1966}, and  recent high redshift surveys have  shown increasing evidence for  an evolution in the power of AGN, their comoving density, and  accretion properties, with a shift in activity from high mass to low mass AGN generally referred to as ``AGN downsizing''  (e.g. \cite{Cowie2003, Merloni2004, Hasinger2005, Kriek2007}). This downsizing in AGN activity appear as part of a more general trend in 
 downsizing of galaxy properties over cosmic times. Star-formation and mass assembly appear to have downsized as well \cite[]{Cowie1996, Cimatti2006}: more massive galaxies were in palace  at earlier times and also  quenched their star-formation earlier than low-mass galaxies, a fact that suggests a physical co-evolution of all galaxy properties and AGN activity as we  discussed in Paper II.

This evolution of AGN power and features seems at odd with a static scenario typical of orientation models where high- and low-power AGN are seen in first approximation as two classes given by different viewing angles. Furthermore, even though orientation clearly affects the properties of AGN and is able to account for many observed features of AGN \cite{Urry1995},  the generality of such models has been challenged in a number of studies  \cite[e.g.,][]{Prestage1983, Moran1992, Malkan1998, Harvanek2001, Chiaberge2002, Tran2003, Celotti2005}. 
Further investigation is therefore warranted to clarify the interplay between host and AGN properties and  the role of orientation and evolution in shaping the properties of active nuclei. 

To this end,  we have carried out a study of a local population of  narrow-line (Seyfert II and LINERs), and broad-line (Seyfert I and Quasars) active nuclei drawn from the Sloan Digital Sky Survey \cite[]{York2000} Second Data Release (SDSS, 2DR). 

In Paper I (Reviglio \& Helfand 2009A) we have presented the multiwavelength database for the narrow-lined population that we have constructed for this study. We discussed the systematics affecting the selection of  active galactic nuclei in redshift surveys, quantifying the incompleteness given by the pollution of SDSS spectra from light of the host and showing that after correction for such effects, $>$50\% of early-type galaxies  show spectroscopic signatures of AGN activity, a fact  suggesting that most, if not all, massive galaxies must undergo an AGN phase in their evolution. Interestingly, we found that, at all host luminosities, emission lines are less pronounced in redder hosts, suggesting that redder galaxies harbor systems with lower (or less efficient) accretion. We further explored this effect in Paper II (Reviglio \& Helfand 2009B), where we showed evidence for an evolution of the spectroscopic signature of AGN, with emission lines becoming less and less remarkable in systems with older stellar populations. This evolution is accompanied by less powerful radio activity and a change in the optical morphology of galaxies which is more substantial in less luminous hosts, in agreement with downsizing scenarios.

In this third paper of the series  we explore the spectroscopic and radio properties of the population of  narrow-line active galactic nuclei discussed in Paper I and  compare their properties with a sample of broad-line AGN and quasars.  We examine the properties of the radio-emitting AGN along with  their spectroscopic features  and further explore the well-established mismatch between the radio, X-ray  and spectroscopic signatures of AGN \cite[]{Martini2002, Best2005, Reviglio2006}.  We study the interplay between the radio emission and spectroscopic features and the physical properties of the supermassive black holes powering them. We  also review critically  unified models, showing major incongruities with the data analyzed in this paper. We discuss the implication of these results.

This paper is organized as follows: in \S \ref{sample} we summarize  the selection criteria and the main properties of the whole sample, while in \S \ref{SMBH} we  derive the masses for the supermassive black holes for the AGN and quasar populations. We go on to show (\S  \ref{host}) that more massive black holes are more likely to power stronger and bigger jets and discuss the role of the environment in such a correlation. In \S \ref{undetected} we explore the properties of the undetected population of radio sources and show a correlation between the median radio power of the spectroscopically selected  population and the mass of the SMBH. In \S \ref{compact} we study the interplay between compactness of radio sources and their photometric, environmental, and spectroscopic properties, showing that the less compact is the radio source, the more unremarkable its lines, lending further support to the idea of a decreasing accretion rate in these systems. In section \S \ref{Unified} we compare the expectations of unified models of AGN activity based on orientation with our sample and show several major inconsistencies. We discuss our results  in \S \ref{Discussion_ch4} and summarize our conclusions in section \ref{Conclusion_ch4}.

\section{The sample}
\label{sample}

We will further analyze here the sample of 151,815 galaxies drawn from the SDSS 2DR  described in Paper I. We refer to that paper for the technical details  concerning the sample selection and the cross correlation with other catalogs. In this paper we will use fully the FIR, radio, X-ray, spectroscopic and morphological information we have cataloged on these objects, as well as the density of their environments  calculated using the three-dimensional density estimator discussed in Paper I and  \cite{Carter01}.

Of these sources, 6\% are detected in the radio band at 1.4 GHz in either the FIRST \cite[]{Becker95} or NVSS \cite[]{Condon1998} Survey, $\sim$ 69\% of which have been classified as AGN on the basis of their spectroscopic and radio properties. The radio luminosity function for these sources is consistent with those from other studies \cite{Sadler2002}, as discussed in Paper I.  In that paper we also showed that by using the FIRST survey's  high-resolution images, the AGN among these radio-emitting galaxies can be  divided in two main categories based on the FWHM size of a gaussian fit to the radio source  :  point-like (unresolved) sources for FWHM$ \le 2.5^{\prime\prime}$ and  jet-like resolved sources for  FWHM$>2.5^{\prime\prime}$. Jet-like sources with typical Fanaroff-Riley morphologies were further classified as FR1 and FR2 sources  \cite[]{Faranoff1974} by visually inspecting all the FIRST fields with multiple radio sources associated with an optical galaxy.

\begin{figure}[h]
\begin{center}
\includegraphics[scale=1.0,angle=0]{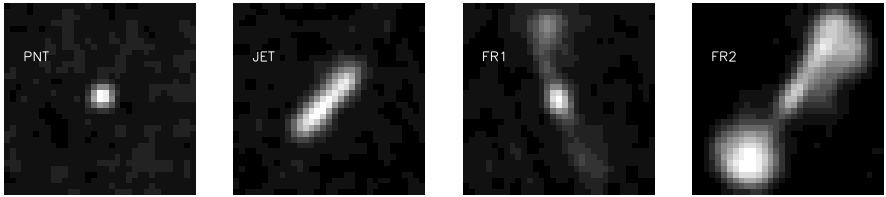}
\end{center}
\caption{\footnotesize \emph{An example for each of the four classes of radio AGN examined in this paper: unresolved point-like sources, jet-like sources, FR1 and FR2 sources.}}
\bigskip
\label{image_sample_tiles}
\end{figure}

In this paper we thus consider four main radio AGN  morphologies: POINT sources, JETs, FR1s and FR2s, including respectively 1935, 2015, 148, and 174 objects. In figure \ref{image_sample_tiles} we show an example of each class. A total of 1739 galaxies had unclassifiable radio emission, since they were not observed in FIRST. 
The median FWHM  of a marginally  resolved jet in the  FIRST survey at the median redshift of the optical survey is $\sim$ 4 kpc -- jets with physical sizes smaller than this would be indistinguishable from point sources. The POINT source class therefore includes intrinsically small, unresolved radio jets.

For the whole optical sample we have adopted the spectroscopic classification  obtained by \cite{Kauffmann2003} and \cite{Brinchmann2004}, based on an improved version of the  standard BPT emission-line ratio diagnostic \cite[]{Baldwin1981}. We also adopted the values of the extinction corrected line-fluxes and  the equivalent widths obtained by those authors, as well as   the extinction-corrected estimates of the stellar masses of the galaxies in the sample. Line fluxes  not corrected for  reddening were obtained directly from the SDSS database, as were measurements of  line widths and information about the velocity dispersion of the bulge-components of the galaxies. 

As discussed in Paper I, dilution and pollution produced by light from the host galaxy's star-forming regions  severely bias the selection of AGN in the SDSS sample, leading to the misclassification of a significant fraction of AGN as passive systems with increasing distance. As shown in Paper I,  the effect of dilution and pollution mainly affects the number counts of early type galaxies, and preferentially affects galaxies with weak emission lines, typically in redder galaxies. All results presented here use our sample in which we  have statistically corrected for this effect.

The Kauffmann et al. 2003 sample does not include broad-line systems, since it was selected to compare the properties of the AGN with their host galaxies. In broad emission line  systems the light of the nucleus often outshines the light of the host and no information about the mass, velocity dispersion, and color of the host can be reliably extracted from the SDSS data. However these systems are of great interests in  understanding the physical properties of different types of active galaxies. For this reason we have selected the whole sample of 1412 broad-line AGN found in the DR2 release at z$<$0.2. According to  standard classifications of quasars,  we divided our sample into 1278 Seyfert Is and 134 quasars requiring quasars to have $M_i<-22$ (Schneider et al.  2005), where M$_i$ is the absolute magnitude of the nucleus in the SDSS i-band.

\begin{figure}[b]
\begin{center}
\includegraphics[scale=0.4,angle=90]{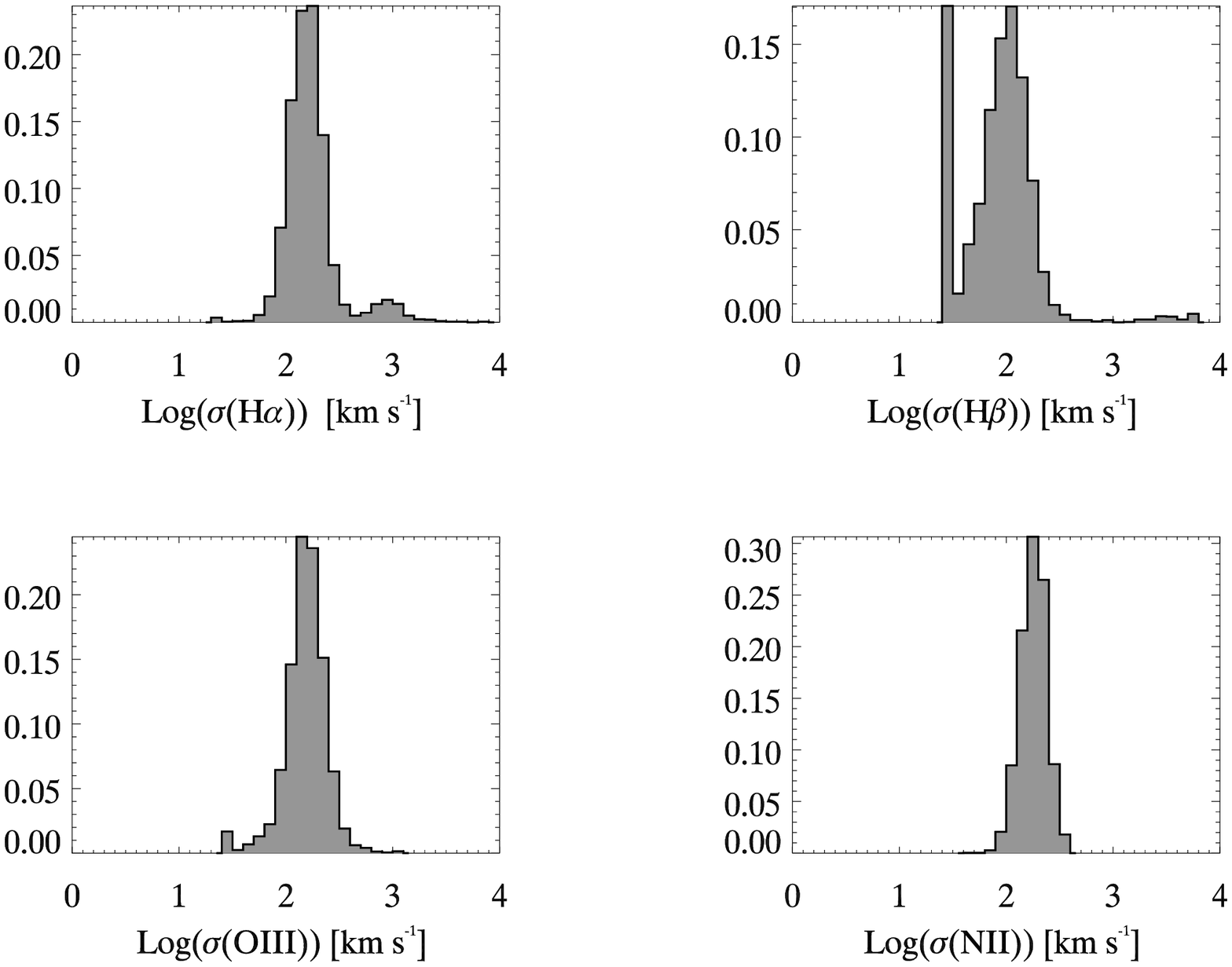}
\end{center}
\caption{\footnotesize \emph{The distribution of line widths for the narrow-line population, selected by Kauffmann et al. (2003)}}
\bigskip
\label{narrow_width_distr}
\end{figure}

\begin{figure}[h]
\begin{center}
\includegraphics[scale=0.4,angle=90]{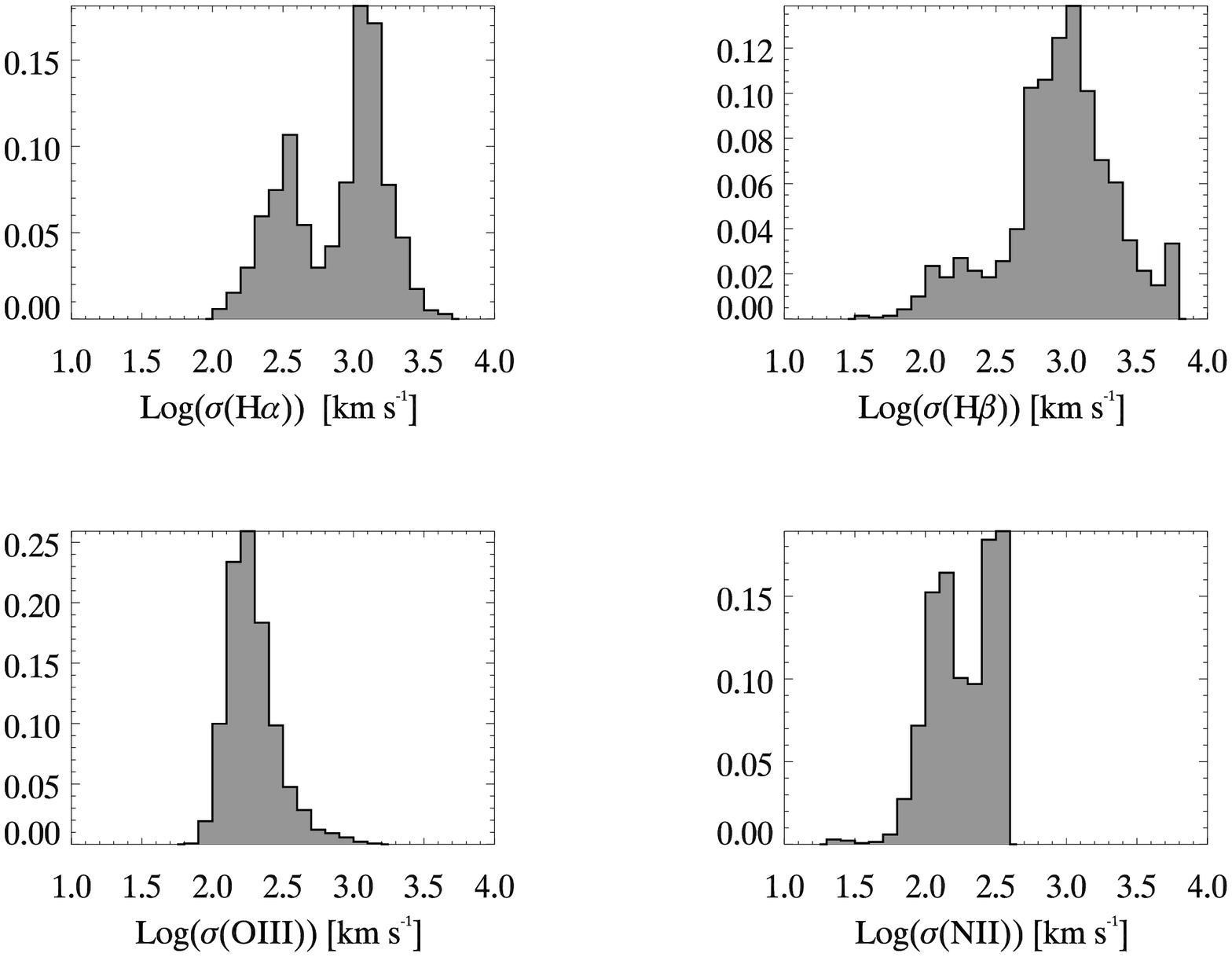}
\end{center}
\caption{\footnotesize \emph{The distribution in line widths for the broad-line population }}
\bigskip
\label{broad_width_distr}
\end{figure}

In order to analyze the relation between line widths and the features of the population of active galactic nuclei, we have downloaded from the SDSS database available line widths  for the sources in both the narrow- and the broad-line samples.
In figures \ref{narrow_width_distr} and \ref{broad_width_distr}  we show the distribution of the $\mathrm{H_\beta}$ line width  for the narrow-line  spectroscopic sample of AGN and for the whole broad-line spectroscopic sample; it is clear  that $\sigma$ $\sim$ 10$^{2.6}\sim 400$ km s$^{-1}$ in H$\beta$ neatly  divides the two classes of AGN, in agreement with the usual notion that a FWHM of 1000 km s$^{-1}$ divides the broad and the narrow population, since  FWHM$=2\sqrt{2\centerdot ln(2)}\sigma$ for a Gaussian fit. A few systems classified as broad-line have either H$\alpha$ or H$\beta$ with FWHM$<$ 1000 km s$^{-1}$, which accounts for the few systems at low FWHM in the histograms of figure \ref{broad_width_distr}. 
A few narrow-line systems show  FWHM$>$ 1000 km s$^{-1}$ (fig. \ref{narrow_width_distr}): we inspected a few of these systems and found that their spectra do now show significant line-broadening, we conclude that these systems have unreliable estimates of their line widths.

For the broad emission line sample we performed the same type of cross correlation  that we did for the narrow-line catalog (cf. Paper I) with the IRAS \cite[]{Moshir90}, ROSAT \cite[]{Voges00}, NVSS \cite[]{Condon1998}, and FIRST \cite[]{Becker95} surveys.
We found 181 matches with FIRST, 44 with IRAS and 474 with ROSAT. All sources detected in FIRST were also detected in NVSS (and vice-versa).
We also visually inspected the radio images and found 13 additional FR2s.
We classified as point-like  106 unresolved sources  and as jets 61  resolved sources with FIRST angular sizes $>$2.5''. One source has a peculiar large jet-like structure, most likely associated with a precessing jet.

\section{Supermassive black holes}
\label{SMBH}

Virial masses of supermassive black holes can be computed from reverberation maps \cite[]{Kaspi2000}. According to these authors,  by assuming that the motion of the gas clouds orbiting  the black hole is  Keplerian, the width of the emission lines  is an indication of the velocity dispersion of the gas $\sigma_g$ and the size of the region producing  the broad-line emission (BLR) of the AGN would be an estimate of the radius.   Under such condition $\mhole=G^{-1}\sigma_g^2~R\sim G^{-1}\sigma_{\lambda}^2~R_{BLR}$. The size of the broad-line emission region can be determined from reverberation maps and has been shown to correlate with the total nuclear luminosity at 5100 \AA~, a fact that allows the estimation of BH masses for large samples of broad-emission line objects.
The masses of supermassive black holes have also been shown to correlate with the blue luminosity of the surrounding hot stellar component \cite[]{Kormendy1992} and with the velocity dispersion of the bulge of the host \cite[]{Ferrarese2000}, providing alternative methods to evaluate SMBH masses in these systems.

We are interested in comparing  the features of different types of AGN with the  masses of their supermassive black holes. For this reason we have computed SMBH masses following different well-tested methods.

For narrow-line objects which typically inhabit spheroids, we have  determined the black hole mass  from the velocity dispersion of the bulge component of the galaxy, $M_\bullet(\sigma)$, following the revised correlation by \cite{Tremaine2002}:

\begin{equation}
\mathrm{log(M_\bullet(\sigma)/\msun)=(4.02\pm 0.44)Log(\sigma/200 km~s^{-1})-(8.13 \pm 0.09)}
\end{equation}

 This relation leads to estimates of $M_\bullet(\sigma)\sim 10^8-10^9 \msun$  for SMBHs in the elliptical galaxies in our sample, in agreement with what is found in the studies of individual systems \cite[e.g.,][]{Kormendy1992}

The \mhole-$\sigma$ relation in galaxies  is known to be a tighter  correlation than the one with the luminosity of the host. However, not all galaxies harboring AGN have measured velocity dispersions. For this reason, and to compare our results with the study of \cite{Best2005}, we have also estimated the masses of the central black holes from the absolute R-band magnitude of their hosts \cite[]{McLure2002}

\begin{equation}
\mathrm{log(M_\bullet(R)/\msun)=(-0.61\pm 0.08)M_R-(5.41 \pm 1.75)}
\label{bh_r}
\end{equation}

\begin{figure}[h]
\begin{center}
\includegraphics[scale=0.4,angle=90]{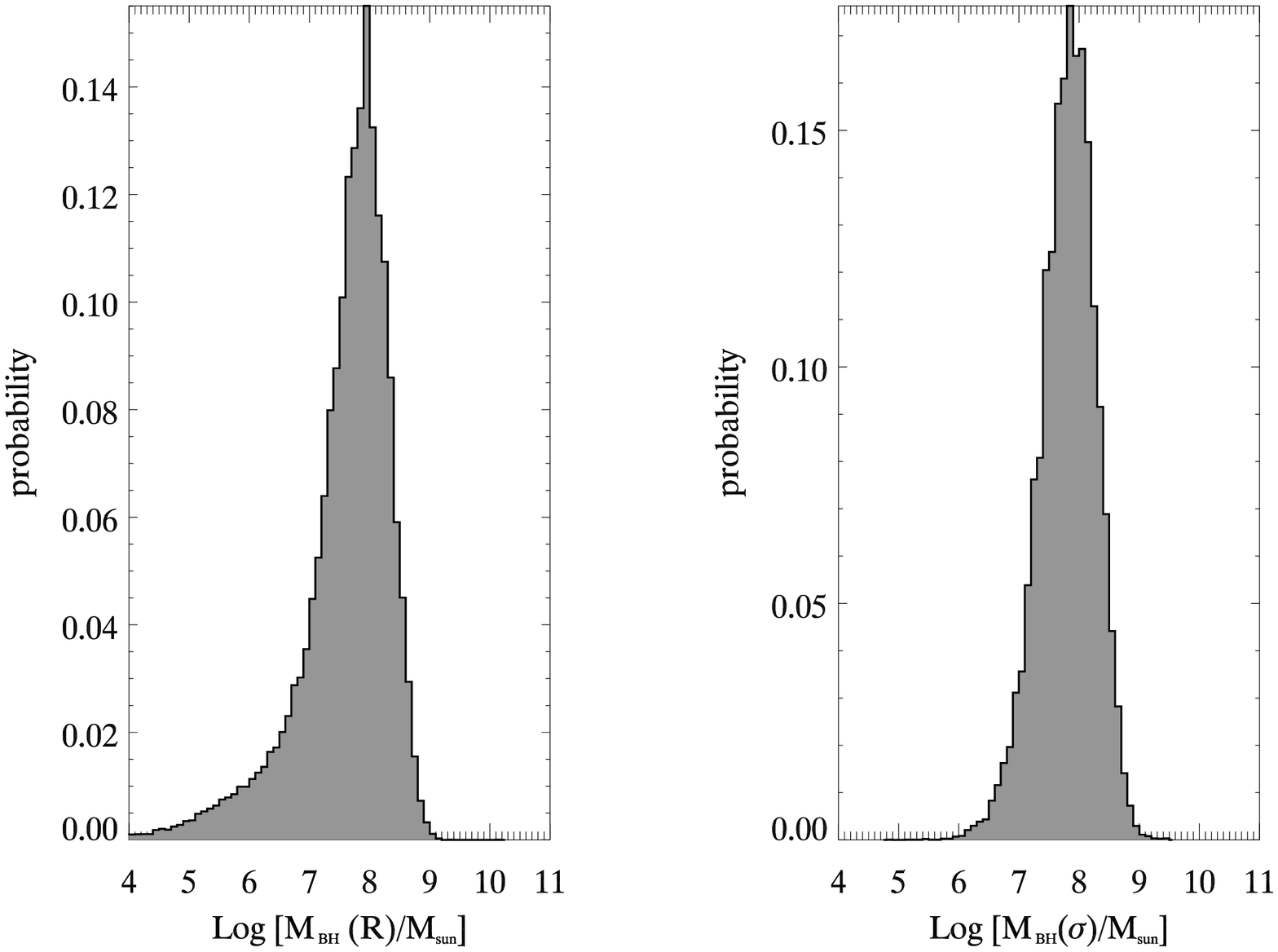}
\end{center}
\caption{\footnotesize \emph{The distribution of  $\mathrm{\mhole(R)}$(left) and  $\mathrm{\mhole(\sigma)}$ for the spectroscopically selected AGN.}}
\bigskip
\label{distribution_bh}
\end{figure}

In figure \ref{distribution_bh} we show the distributions of SMBH masses for the whole spectroscopic population of narrow-line AGN, corrected for dilution and contamination as discussed in Paper I, obtained with the two methods outlined above.

The plots show that the two estimates are largely consistent, but the absolute magnitude method  produces a long tail at the low-mass end. The skewness of the distribution suggests that this is not an effect of the large uncertainty in the mass estimate, since this  would give a symmetric broadening of the distribution. One might argue that since an absolute R magnitude can be calculated for all systems in the survey, including bulgeless late-type galaxies, these might account for the low-mass tail. However, the systems in the low-mass tail are spectroscopically classified as AGN and therefore \emph{must} harbor a massive black hole. We note that for systems with $\mathrm{\mhole(\sigma) < 10^{6.5} M}\odot$ the median concentration parameter  $C=R_{50}/R_{90}$ given by the ratio of  $\mathrm{R_{50}}$ (the radius $\mathrm{R_{50}}$ within which 50\% of the light of the galaxy is emitted) and  $\mathrm{R_{90}}$ (the radius $\mathrm{R_{90}}$  within which 90\% of the light of the galaxy is emitted)  is  0.4310. For comparison, late-type galaxies as defined by \cite{Strateva2001} have  C=0.447, while early-type systems have  C=0.351. This shows that the tail arises because of  late-type galaxies with small/unremarkable bulge componenent harboring low-mass SMBH. This is supported by the difference in the $\mathrm{p=atan(L_{deV}/L_{exp})}$ parameter given by the arctangent of the ratio of the likelihood that a de-Vaucouler's and  an exponential disk profile fit the host surface brightness:   the median $p$ for the  $\mathrm{\mhole(\sigma) \gtrsim 10^{6.5} M}\odot$ is  1.457, typical of bulgy systems,  while the median p for the    $\mathrm{\mhole(\sigma) \lesssim 10^{6.5} M}\odot$ population is 0.148, a factor 10 lower and   typical of disk galaxies. 

For the broad-line population, we  determined virial BH masses  $M_\bullet(H\beta)$ from the H$\beta$ line profile and the  total $L_{5100}$ luminosity of the nucleus  following the revised relation by \cite{Greene2005}:

\begin{equation}
\mhole(H\beta) = (4.4 \pm 0.2) \times 10^6 \left(\frac{\lf}{10^{44}~{\rm erg~s^{-1}}} \right)^{0.64 \pm 0.02}\left( \frac{\fwhb}{10^{3}~\mathrm{km~s^{-1}}} \right)^2 \msun .
\end{equation}

The light of the AGN largely dominates the emission in all the SDSS bands for broad- emission-line objects. Therefore the total luminosity $L_{5100}=\lambda_{5100} \centerdot \mathcal{L}_{5100}$ of a nucleus can be determined by scaling the SDSS  R-band specific luminosity $\mathcal{L}_{6165}$ for the object to the specific luminosity at 5100 \AA~,  $\mathcal{L}_{5100}$. The scaling is done using a power-law spectrum $\nu^{-\alpha}$ typical for broad-line AGN continuum with  power spectrum index $\alpha=0.5$: $\mathcal{L}_{5100}=(6165/5100)^{(1-\alpha)}\mathcal{L}_{6165}$.
 The best choice of SDSS magnitude system for objects in which the point-like   nuclear emission dominates the photometric images, such as in quasars and Seyfert Is, is the point-spread-function magnitude (PSF-mag).
Therefore  R-band specific luminosities have been computed from the SDSS R-band  PSF-magnitudes. The results do not change  significantly  if instead of using the R-band luminosity to obtain $\mathcal{L}_{5100}$ we use the U-band luminosity.  We also compared our estimates for the SMBH  masses of Seyfert I nuclei and quasars to those obtained by \cite{Sikora2007} who used a different sample and a different method of SMBH mass estimation;  we found very good agreement.

\begin{figure}[h]
\begin{center}
\includegraphics[scale=0.4,angle=90]{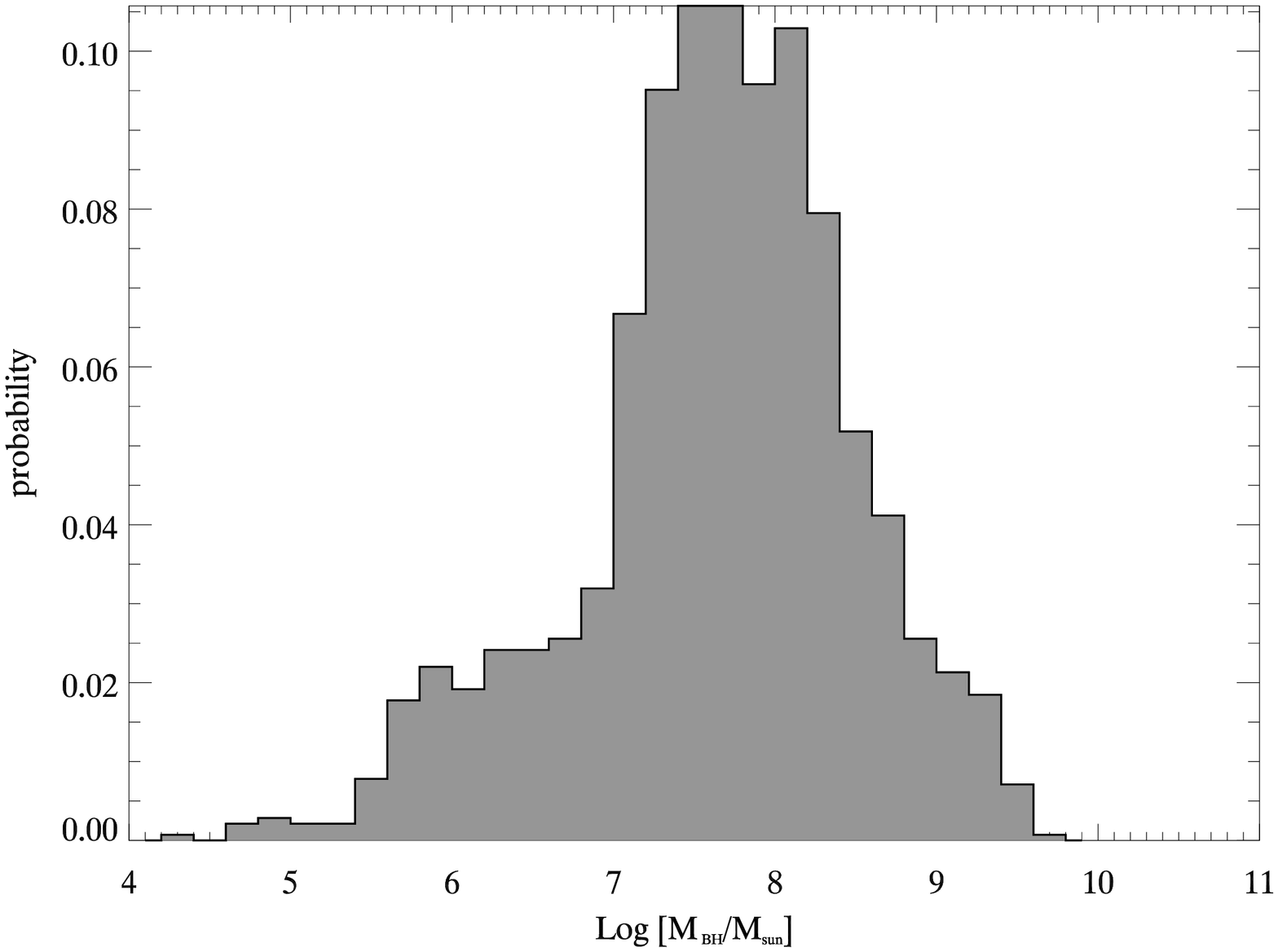}
\end{center}
\caption{\footnotesize\emph{The distribution of  $\mathrm{\mhole}$  for broad emission-line AGN.}}
\bigskip
\label{distribution_bh_broad}
\end{figure}

In figure \ref{distribution_bh_broad} we show the distribution of the supermassive black hole masses for the broad-line spectroscopic  population of AGN. The distributions of SMBH masses for the narrow and broad line populations span the same range of values and peak at fairly similar values, supporting the idea that it is not the mass of the SMBH that drives the difference between the two classes.

\begin{figure}[h]
\begin{center}
\includegraphics[scale=0.4,angle=90]{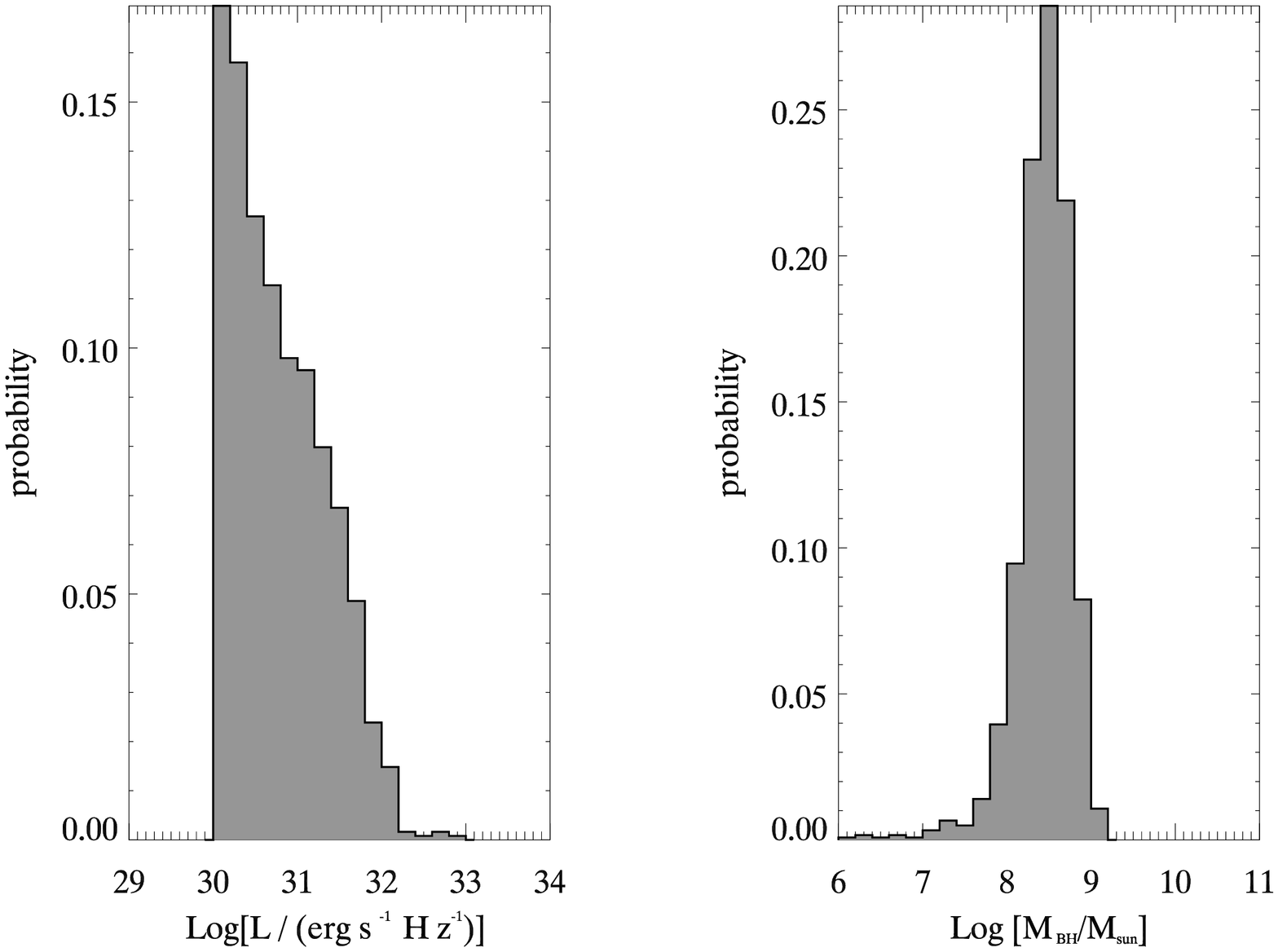}
\end{center}
\caption{\footnotesize \emph{The distribution in luminosity (left) and central black hole mass $\mathrm{\mhole(R)}$(right) of the final sample of 1215 radio sources classified as jets in the narrow emission line systems.  $\mathrm{\mhole(R)}$ values were derived from the galaxies' R  magnitudes}}
\bigskip
\label{narrow_jet_distr}
\end{figure}

In figure \ref{narrow_jet_distr}  we show the distribution in luminosity for 1215 radio jets more luminous than $10^{30} \lradio$ for  the narrow- emission-line population, along with the distribution of their central black holes masses.
The distribution in luminosity of the jets falls off rapidly towards higher luminosities, while the distribution in masses of the black-holes peaks at intermediate black hole masses.

\begin{figure}[h]
\begin{center}
\includegraphics[scale=0.4,angle=90]{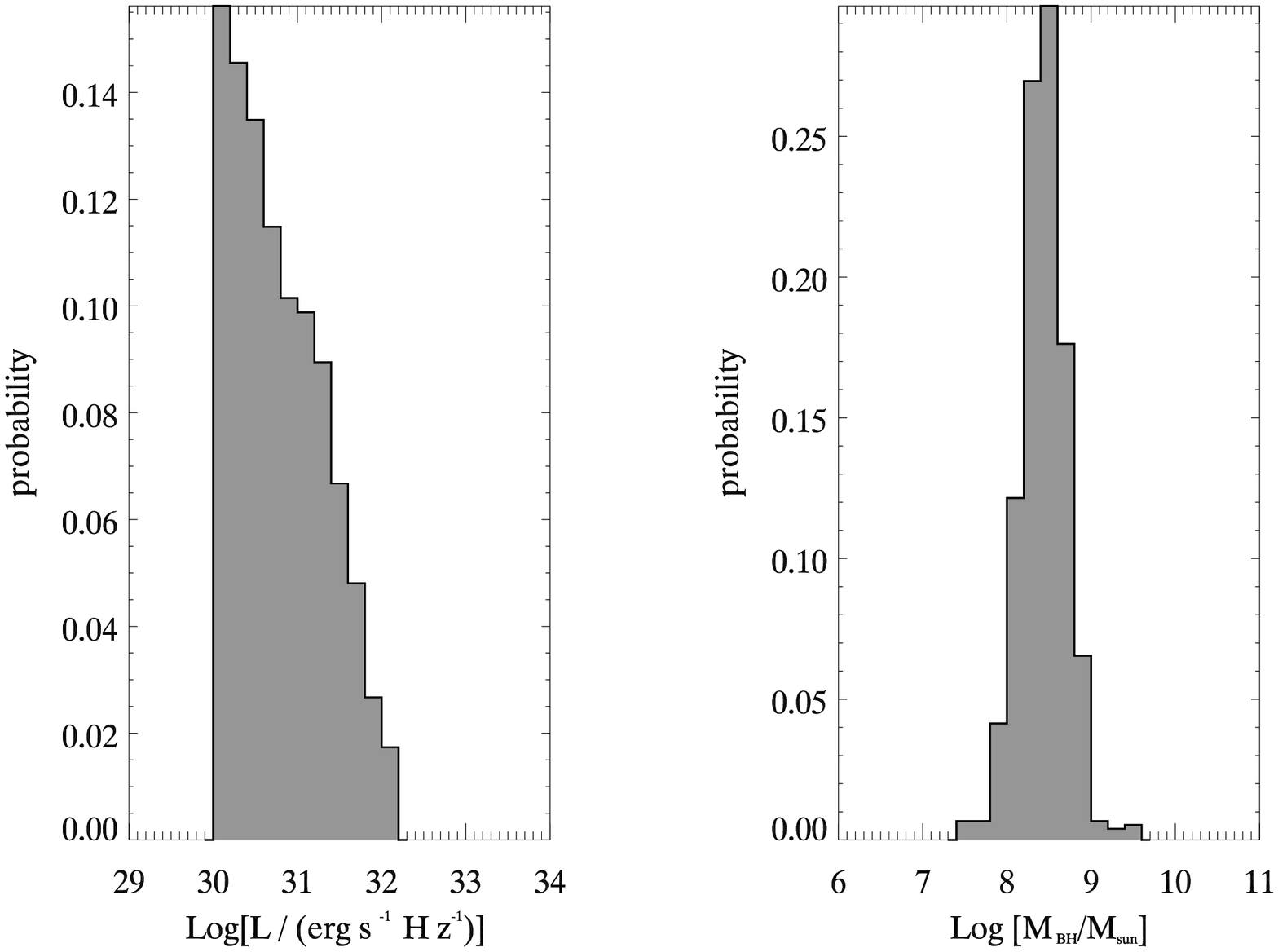}
\end{center}
\caption{\footnotesize \emph{The distribution in luminosity (left) and central black hole mass $\mathrm{\mhole(\sigma)}$(right) for a subsample of 749 systems in narrow emission line sample classified as jets for which bulge velocity dispersions were available, in narrow emission line sample.}}
\bigskip
\label{narrow_jet_distr_sigma}
\end{figure}

For comparison, in figure \ref{narrow_jet_distr_sigma}  we show the distribution in luminosity for a sub-sample of 749  radio jets more luminous than $10^{30} \lradio$ which have black hole masses measured from bulge velocity dispersions $\mathrm{\mhole(\sigma)}$, as well as the distribution of their $\mathrm{\mhole(\sigma)}$. We note that the distribution in luminosity and in black hole mass are very similar to those  determined using the R-magnitude method. This demonstrates that the choice of the black hole mass estimator does not significantly affect the statistical properties of the jet sample. It also shows that the distribution of the jets is not significantly affected by the optical selection of the sample: drawing a sub-sample from the original one with different optical properties does not change the shape of the distribution significantly.

\begin{figure}[h]
\begin{center}
\includegraphics[scale=0.4,angle=90]{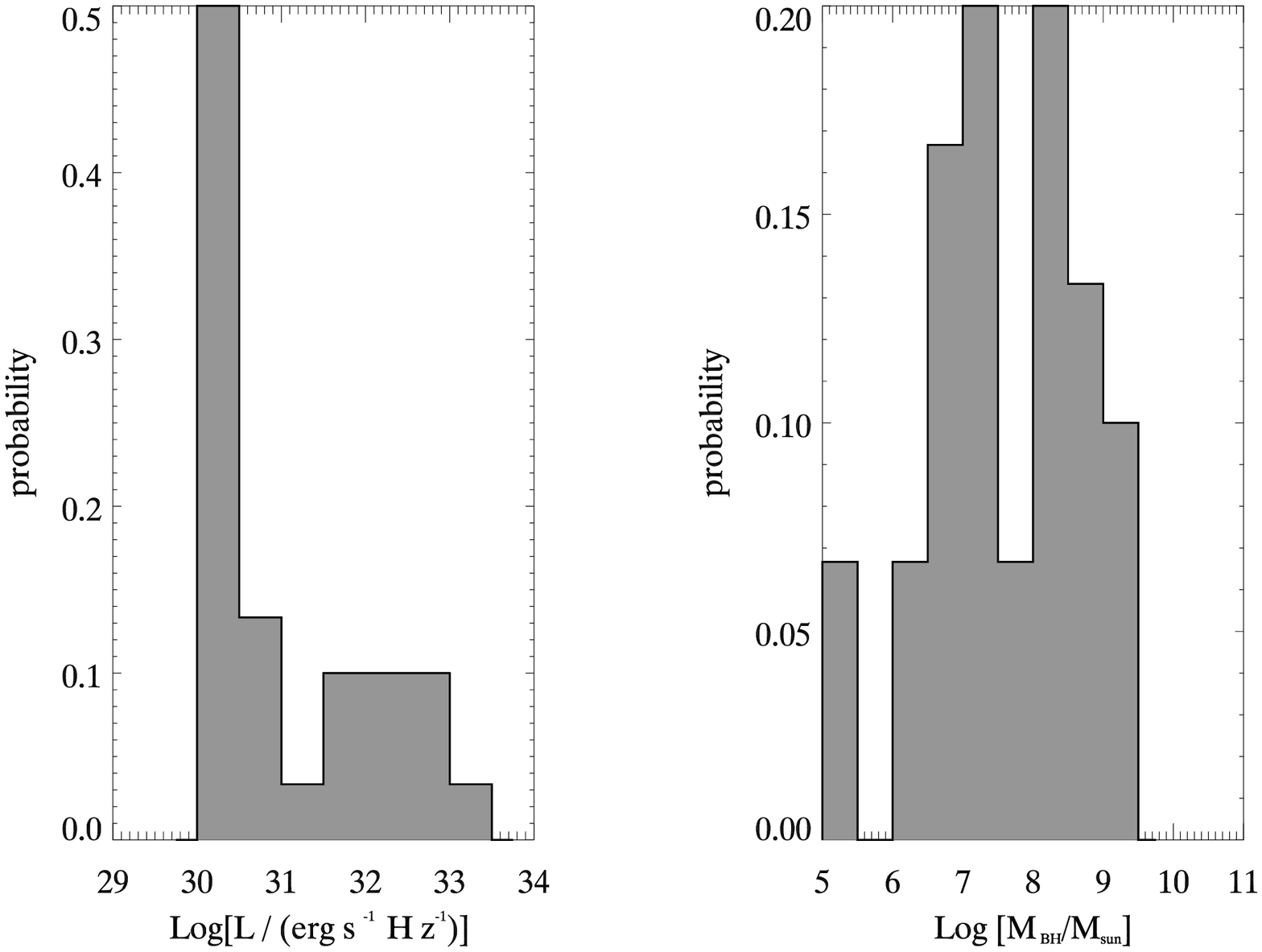}
\end{center}
\caption{\footnotesize \emph{The distribution in luminosity (left) and central black hole mass(right) for radio sources classified as jets in broad-emission-line systems.}}
\bigskip
\label{broad_jet_distr}
\end{figure}

In figure \ref{broad_jet_distr}  we show the distribution in luminosity of radio jets more luminous than $10^{30} \lradio$ for  the broad-  emission-line population and the distribution of their central black hole masses. The figure suggests a bimodal distribution  in both radio luminosity and black hole mass, suggesting a link between the radio luminosity and the mass of the SMBH. We will  investigate this further in the next section.

\section{Dependence of radio  AGN activity on the SMBH and the environment}
\label{host}

There are primarily two questions we wish to answer: do more luminous galaxies host more luminous radio AGN? And, do more luminous galaxies host bigger jets?
These questions cannot be answered cleanly owing to selection effects arising from the flux-limited nature of the optical and radio surveys, which favors the selection of brighter hosts and brighter radio sources at higher redshifts.

We would be satisfied by answering two other questions: are radio AGN with luminosity above a certain threshold $L_t$ more likely to be found in more luminous host galaxies? And, are radio jets bigger than a certain size $S_t$ more likely to be found in more luminous host galaxies?

In order to answer these questions we used a chain of eight volume-limited samples with widths of 0.5 in absolute R magnitude over the interval $-23.5<R<19.5$.  For each of these classes we determine the fraction of the members of the class  which have an AGN with \emph{radio} luminosity greater than $L_t$, where $L_t$ is a radio  luminosity detectable throughout the whole volume. As above, we use $ L_t$=$10^{30}$ \lradio; if an AGN is brighter than $L_t$, it would be detected in FIRST at all redshifts below z=0.2, the cut-off adopted for our optical sample. More of these sources would be detected at higher redshifts, because of the geometry of the surveys, since at higher redshift larger volumes are sampled and therefore one has greater chance to select brighter objects. However, this volume effect is removed by normalizing this number of brighter-than-$L_t$ AGN  to the number of optical galaxies in the class.

For each optical absolute magnitude class we then determine the median SMBH mass and plot the fraction of sources brighter than $ L_t$=10$^{30}$ \lradio  as a function  of this median SMBH masses. The advantage of this method is that one samples several orders of magnitude in luminosity for the host galaxies ( and therefore several orders in SMBH masses, given the scaling relation between R-band luminosity of a host galaxy  and its SMBH mass). The fact that low-luminosity AGN are not seen at higher redshift is accounted for by restricting the analysis to the sub-sample of AGN that would be observable at all redshifts throughout the survey. Nothing can be inferred for AGN with luminosities lower than $L_t$. We will extend this analysis to low-luminosity objects in the next section by means of another technique.

\begin{figure}[h]
\begin{center}
\includegraphics[scale=0.5,angle=90]{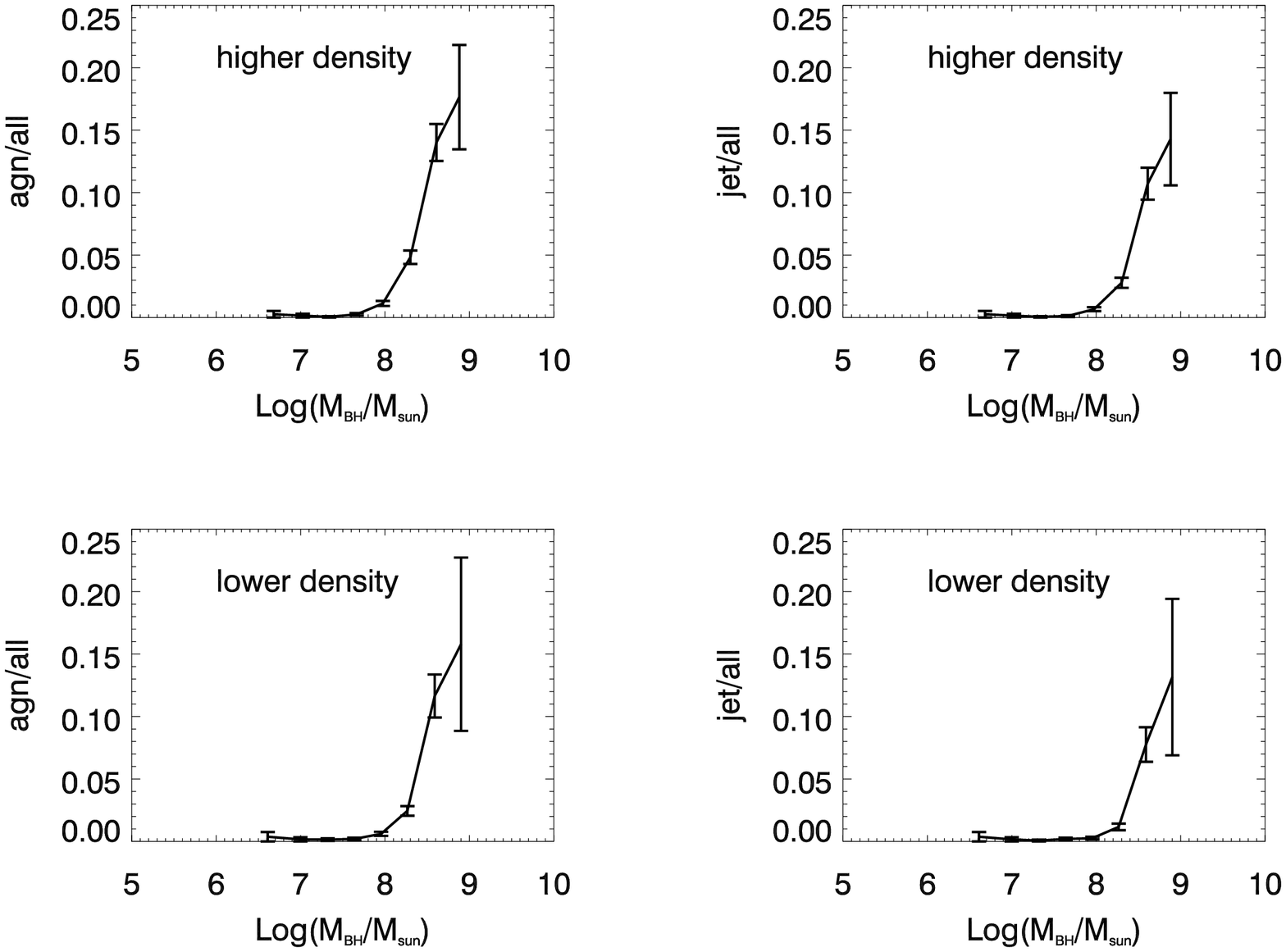}
\end{center}
\caption{\footnotesize \emph{Left: the fractional abundance of radio AGN with luminosity above $L_t$ as a function of baryonic mass of the host. A steady increase up to two order of magnitudes in the fraction of bright radio AGN is found. Right: when jet radio sizes  are considered, we find that the fractional abundance of large jets strongly increases with increasing mass (or equivalently luminosity) of the hosts.}}
\bigskip
\label{jets_fractions_bh}
\end{figure}

Since we wish to break the degeneracy between physical properties of the host and its environment which results mainly from the density-morphology relation, we have divided our sample in two narrow classes of densities: lower densities (0.5$<$Log($\rho$)$<$1.0) and higher densities (1.0$<$Log($\rho$)$<$1.5).
The results of this analysis are shown in the left column of fig.\ref{jets_fractions_bh}. All errors are calculated assuming Poisson statistics. 
Galaxy luminosities are proportional to their stellar masses, so the plots can be interpreted either  in terms of host luminosity or its baryonic mass. 
In bins below $\mathrm{Log(\mhole(R)/M_{\odot})=8.0}$, the fraction of AGN with luminosity above $L_t$ is negligible. With increasing luminosity this fraction increases by two order of magnitudes. For galaxies with SMBH masses higher than $10^{8.5} M_\odot$ more than 15\%  have a bright radio AGN. Similar trends are found if, instead of considering the  whole population of AGN, we restrict the analysis just to jet population. More massive SMBHs power stronger radio sources, in agreement with models of jet formation \cite[e.g.,][]{Blandford1977}.

The analysis of jet sizes was carried out in a similar way.
For each of the classes of optical luminosity, we evaluated the fraction of those galaxies with jets larger than a size $S_t$ which would be resolved throughout the whole survey. We chose  S$_t$=7.5 kpc, for which jets would be resolved up to z=0.2.
Once again we removed the volume effect by normalizing the number of jets larger than $S_t$ in a certain optical luminosity bin by the number of galaxies in that bin. To avoid problems with low-luminosity jets  that might have been missed in the FIRST Survey because of its flux-density limit,  we restricted our analysis to jets with  $ L_t$=30 \lradio. We then plotted the fraction of galaxies with jets larger than a size $S_t$ against the median SMBH mass for each class. Physical sizes for our jets have been calculated by multiplying the angular size of our source by the angular distance $D_A$ which relates to the luminosity distance $D_L$ as $D_A=D_L/(1+z)^2$. 

The results are shown in the right column of fig.\ref{jets_fractions_bh}. In bins below $log(M/M_{\odot})=7.5$ the fraction of AGN with jet size  above $S_t$ is negligible. With increasing luminosity the fraction increases by two order of magnitudes; more than 15\% of galaxies with masses above 8.5 have a large radio jet. We cannot correct for the orientation of the jets. However, since jet orientation is  not dependent on the host luminosity or the SMBH mass, the trends found cannot be due to  such effects. More massive SMBHs have a much greater chance of powering radio jets of larger size.

A similar increase in the fractional abundance of high-power radio AGN and jets with SMBH mass is seen in both high and low density environments; the same is true for the increase in the fractional abundance of larger jets with increasing SMBH mass. This means that, to first approximation, the physical properties of the host determine the size and  power of radio AGN in radio galaxies, not the environment in which the jet propagates or the effects of interaction with neighboring galaxies or the intergalactic medium. In particular, the existence of large and powerful jets depends on the mass of the SMBH (or equivalently on the mass of the host galaxy). This is in agreement with simulations of jets propagating in an external medium. \cite{Zanni2003} have followed  the propagation of jets in a stratified external medium and concluded that two regimes exist: one phase in which the jet cocoon is highly over-pressured with respect to the external medium and drives a strong shock into the ambient gas, and  a second phase in which the cocoon pressure decays and reaches balance with the external value, making the shock become weak.  The separation between two regimes depends essentially on the jet kinetic power; the transition occurs at larger distances from the galactic nucleus for high kinetic-power jets and at lower distances for low kinetic-power jets (cf. also Ferrari, 2006). Therefore one would expect stronger radio  jets (which generally have high kinetic power as evinced in their high Mach numbers) to have on average larger radio sizes, since it is the first phase that sets the observed size of a radio jet. Re-acceleration of electrons or positrons in a weak shock is much less effective than in strong shocks \cite[]{Fermi1949}; in addition,  the magnetic fields are much lower far from the galaxy, leading to much weaker synchrotron emission in the second phase.

\begin{figure}
\begin{center}
\includegraphics[scale=0.5,angle=90]{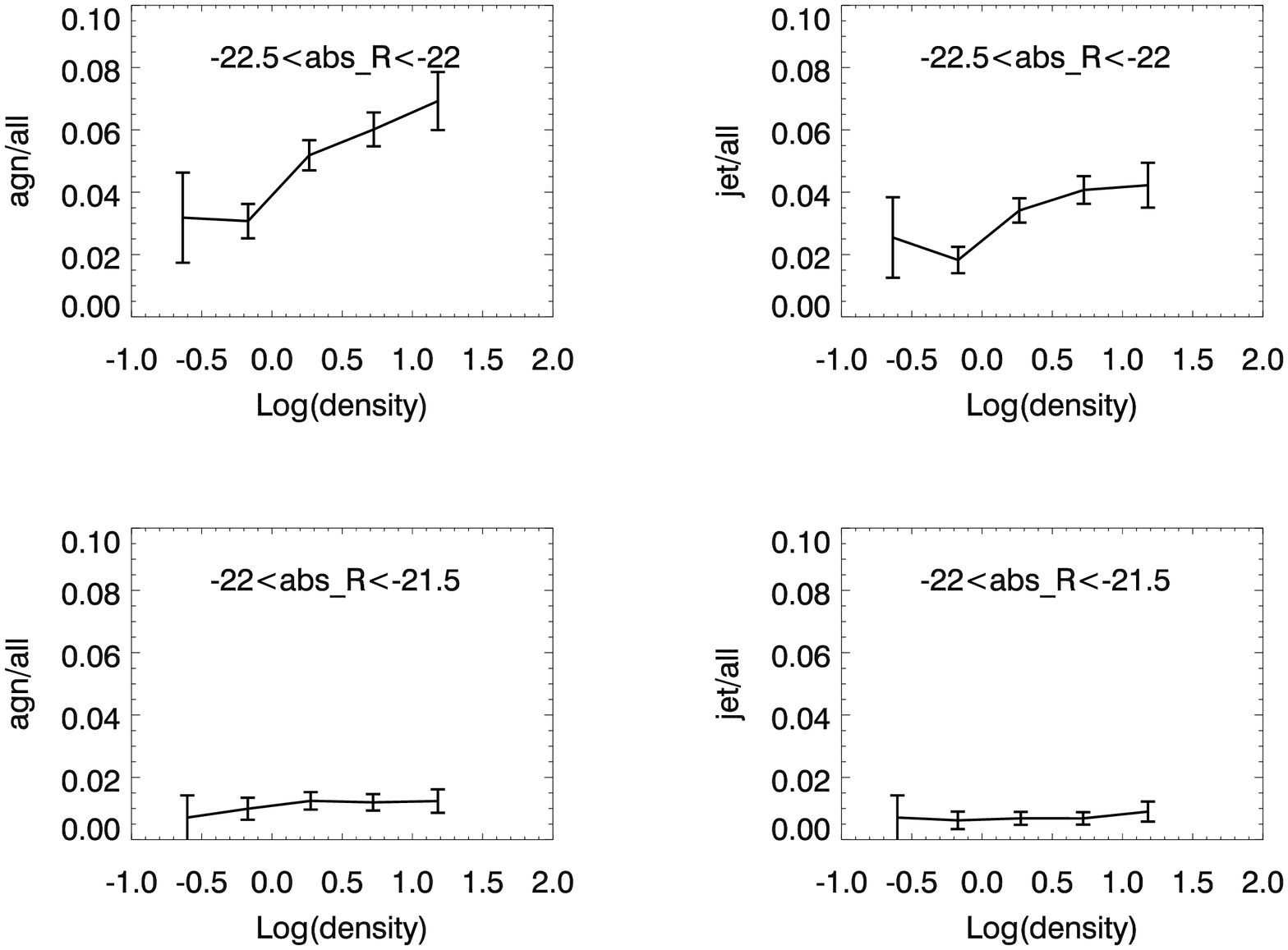}
\end{center}
\caption{\footnotesize \emph{Left: the fractional abundance of radio AGN with luminosity above $L_t$ as a function of density of the host environment. A mild  increase of a factor of two is found for luminous galaxies in denser environments. Right: when jet radio sizes  are considered we find that the change of powering large jets is higher in massive galaxies residing in denser environments.}}
\bigskip
\label{jets_fractions_dens}
\end{figure}

In order to determine if the environment is correlated with the existence of larger and more luminous jets   we fixed two classes in absolute magnitude (-22.5$<$R$<$-22.0 and -22.0$<$R$<$-21.5) and  plotted  the fraction of galaxies with  powerful (L$>$ 10$^{30}$ \lradio) and large (S$_t>$7.5 kpc) jets with different environmental density (figure \ref{jets_fractions_dens}). We find a modest  increase of nearly a factor two  between low- and high- density environments in the likelihood of finding stronger radio sources and bigger radio jets, although this is evident only for the higher luminosity galaxy sample.

It appears therefore that even though the role of the environment is limited compared to the role of the mass of the SMBH in triggering bigger and more powerful radio jets, its effect is not negligible. Since denser galaxy environments are generally associated with denser intergalactic media the pressure of which should tend to confine the jet, not make it bigger, we suggest that the effect of the environment on powering jets may be given by  the interactions of the radio galaxy with their neighbors which are more frequent in denser environments. Such interactions could disturb the gas in the  host, making it easier for it to fall in the inner regions of the galaxy and feed the active nucleus.  If the accretion is  higher, stronger radio jets are expected in current models of jet formation (cf. Meier, 2001)

\section{The undetected population}
\label{undetected}

Only 6\% of the galaxies in our sample have detected radio emission: the vast majority of active galaxies are not detected  at the current survey sensitivity threshold:  either  the flux-limited nature of the radio survey is responsible or  there may exist a population of active galaxies actually lacking radio emission. We employ radio image stacking to evaluate the median properties for the whole radio AGN population, detected and undetected. This allows us  assess whether more massive galaxies power more powerful radio AGN or not, a question that we have left unanswered in the previous section.

We have  used the procedure described in White et al. (2006) for stacking FIRST  fields with and without radio emission  and evaluate   median properties of the resulting radio image. Since we are evaluating \emph{median} images, including the actual detections does not bias the result; on the contrary, it allows us to explore the whole radio luminosity distribution of the sample.  It is worth stressing however that the undetected population represents $\sim$94\% of the sample.

%\begin{figure}[h]
%\begin{center}
%\includegraphics[scale=0.5]{/scratch/reviglio/huitzil/starform/sloan/analysis/paper_plots/tiled_images.eps}
%\end{center}
%\caption{\footnotesize \emph{From top to bottom: stacked images for 8 volume-limited samples with increasing absolute magnitude of spectroscopically classified star-forming galaxies (left column) and AGN  (right column). }}
%\bigskip
%\label{tiled_images}
%\end{figure}

We limit our analysis to the usual chain of volume-limited samples with R between -23.5 and -19.5  for optical completeness and we divide our sample according to its spectroscopic properties.  For each class of luminosity we evaluated the median SMBH mass  and the median radio luminosity. Median radio luminosities are calculated by fitting 2-D gaussians to the maps obtained by stacking fields  in luminosity space.

\begin{figure}[b]
\begin{center}
\includegraphics[scale=0.4, angle=90]{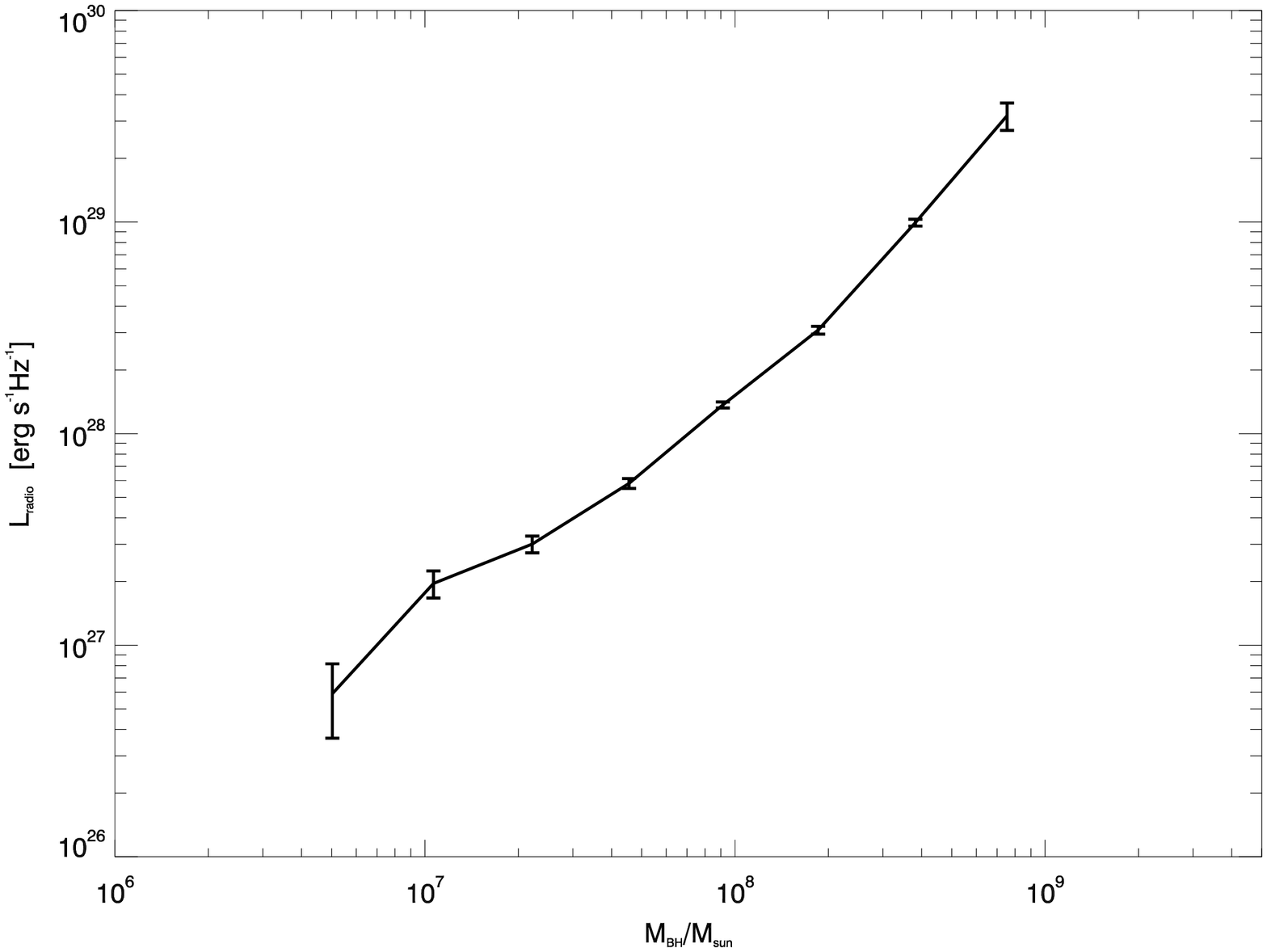}
\end{center}
\caption{\footnotesize \emph{Trends of the median radio luminosity as a function of  SMBH  mass for stacked images of spectroscopically classified AGN. The average luminosity of radio jets increases with increasing SMBH mass ( or, equivalently, luminosity  of the host). The error bars represent the luminosity of the first and third quartile stacked images.}}
\bigskip
\label{stack_trend_lum_mass}
\end{figure}

We find that an increase in the mass of the SMBH (or equivalently the absolute magnitude of the host) is accompanied by  a strong increase in the median luminosity of the radio AGN ( cf. fig. \ref{stack_trend_lum_mass}). This trend might even be underestimated since, as shown above,  the most massive galaxies tend to have large jets and  with the technique used here we are only able to evaluate the central luminosity of the radio AGN;  any luminosity in the lobes is missed. This allows to conclude that more luminous central components of radio AGN are associated with more massive SMBH: this is in qualitative agreement with expectations of jet production theories \cite[e.g.,][]{Meier2001}. This result is in disagreement with the claim by Best et al. 2005 that the distribution in radio luminosity of active nuclei is independent of the SMBH mass.

\section{Compact vs. extended radio sources}
\label{compact}

Galaxies are known to show a variety of radio  structures.
While some galaxies show point-like radio sources, other show radio jets that might stretch out well beyond the optical host. The mechanism(s) leading to such differences is not understood. The alignment of jets postulated by orientation models has been challenged by several studies  \cite[e.g.][]{Prestage1983, Harvanek2001} on the basis of the diversity in environment where these sources are found. Alternative explanation contemplate frustration of the jet by the interstellar medium\cite[]{vanBreugel1984} or an evolutionary scenario \cite[e.g.,][]{Odea1998}. 

In \S \ref{host} we have shown a strong correlation of jet size with the mass of the host galaxy; however, more massive galaxies are also bigger. In this section we compare the size of the jet with the size of the host galaxy, trying to evaluate if the compactness of the jets correlates with  host and/or environmental properties.

First, we note that point-like structures in the narrow-line population can be physically small, unresolved jets or larger jets that appear smaller either because of their orientation with respect to the observer or because of their distance, since the physical size resolved  throughout the whole survey up to $z<0.2$ is 7.5 kpc.  We note that  point-like sources inhabit preferentially lower density environments than jet-like structures: their respective  average environmental density is 2.87 galaxies/Mpc$^3$ and 3.50 galaxies/Mpc$^3$; in addition their respective median radio luminosities  are different, with point-like sources average luminosity of 10$^{29.9}$ \lradio and jet average luminosity of  10$^{30.4}$ \lradio. Since orientation does not depend on the environmental density and the luminosity of the point-like sources is expected to be higher rather than lower than the luminosity of the jet population if beaming is present,  this argues against the hypothesis that orientation effects mostly account for the unresolved population.  
As far as the possibility that point-sources are the distant unresolved sub-sample of the standard jet population, we note that  this is not the case, since they tend to reside at somewhat lower redshifts than jets: their median redshift is $\sim$ 0.11 , while the median redshift of jets is $\sim$ 0.14. 
Therefore point-like sources  must be, for the most part, \emph{physically} different from larger jets, as suggested by their lower luminosities. 

For the SDSS sample, a measure of the optical size of the galaxy is given by the  Petrosian radius (Blanton et al. 2001).
We therefore consider three main classes of objects: point-like (unresolved) sources, compact sources with jet size  $S_j$ smaller than the R-band Petrosian radius $P_r$ of their host and extended jets with  $S_j>P_r$. 

We expect some degree of contamination in the second class from larger jets oriented close to the line of sight. We will show that our results are not significantly affected by this problem. We also expect some degree of contamination in  the first class, given by large jets that might become unresolved at the highest distances. We note however that the high-resolution of the FIRST Survey limits this problem significantly. Only jet smaller than 7.5 kpc are unresolved at $z \sim 0.2$

We evaluate the median  absolute R magnitude for the three samples for all sources with $z < 0.2$: $\ovr_p$ for the point-like class median, $\ovr_c$  for the compact class and $\ovr_e$ for the extended class.
We find a increasing trend in median luminosity with decreasing compactness of the radio source:  $\ovr_p$=-21.98,  $\ovr_c$=-22.20,  $\ovr_e$=-22.41 with median absolute deviations of 0.47, 0.41, 0.32, showing that the spread of the distributions in absolute magnitude for the three classes decreases with decreasing compactness, a likely effect of the contamination of the hosts of compact sources by the hosts of extended sources seen at an angle, which tend to be more luminous and therefore add to the spread in luminosity in these classes.

In order to establish if  the decreasing trend in the median luminosity is statistically significant, we chose  to apply  a chain of Wilcoxon (Mann-Whitney) tests to pairs of the  median values  $\ovr_p$, $\ovr_c$,  $\ovr_e$,  in order to establish if, given the variance of the distributions, their differences are significant. First we compared  $\ovr_p$  with  $\ovr_c$ and found that the probability that the difference in the two median values is not significant is of order $10^{-5}$,  establishing that $\ovr_p > \ovr_c$. Similarly we applied the test to  $\ovr_c$ and $\ovr_e$ and established that  $\ovr_c > \ovr_e$  at the same confidence level. Thus,  $\ovr_p> \ovr_c > \ovr_e$ and the decrease in absolute magnitude (or \mhole) is significant.

In order to ensure that blending of these classes does not significantly affect this conclusion we evaluated the median absolute magnitude for the three classes applying more restrictive cuts: we required that jets in the compact class  have sizes  significantly smaller than the Petrosian radius of the galaxy    ($S_j<0.7\times P_r$), therefore ensuring a lower chance of mistakenly including large jets oriented along the line of sight in this class;  we also limited in redshift the population of  point-like sources by selecting only sources at  redshifts $z< 0.1$, therefore ensuring a lower contamination in this class from larger unresolved jets that would belong to the compact class.  We find $\ovr_p$=-21.45,  $\ovr_c$=-22.10,  $\ovr_e$=-22.41: the trend is still present, and blending does not appear to introduce a significant bias.

All  three classes of objects are found at all optical luminosities, but their median optical luminosities appear to decrease gradually for more and more compact sources. This also means that the average mass of their SMBH decreases with increasing compactness of the radio source. Indeed, we find  that the median SMBH masses for the three classes are $10^{8.13}$, $10^{8.32}$, and $10^{8.46}$ M$_\odot$.
There are no selection effects that would produce correlation between the compactness of an AGN and the luminosity of its host. Point-like sources are seen both in high- and low-luminosity systems. For extended sources there might be some problem with the fact  the FIRST survey's high-resolution tends to resolve out part of the flux of extended sources such that  the true size of some jets might be underestimated. This problem however does not preferentially happen in either high or low optical luminosity systems.

We performed the same type of analysis considering the U-R colors of the hosts and found  respectively 2.87, 3.03 and 3.14 mag, and a significant correlation from a chain of Wilcoxon (Mann-Whitney) tests, showing that more compact sources are found in bluer systems. 

If we compare the prominence of the bulges along this sequence  we find  a small, yet significant  increase  in the bulge to disk ratio, with p=arctan($L_{deV}/L_{exp}$): we find respectively  1.478, 1.505, and  1.519. The MAD for the distribution are respectively 0.028, 0.039, and 0.070,  and a  Wilcoxon test confirms the existence of a significant trend. Using the concentration of the light profile, brings to the same conclusion. This is also  in agreement with the fact that the host colors are redder for less compact jets.

For the median densities of the environment surrounding these sources, we find respectively 2.88, 3.40, 5.00 galaxies/Mpc$^3$, and  positive correlation from a chain of Wilcoxon (Mann-Whitney) tests, showing that more compact sources are found in less dense environments. Interestingly, we find that our low-luminosity point-like sources are found in lower density environments, similarly to their high-luminosity counterparts studied by \cite{Prestage1983}.

Therefore, when the size of the jet is compared to the size of the host, we find that, on average, less compact jets are found in redder, more luminous, more bulgy, early-type galaxies, which typically reside in denser environments, and that the  transition from compact to extended jets is \emph{smooth} along this portion of the Hubble Sequence.

We compared the spectral properties of the three classes of radio  compactness. We selected as usual a small range in absolute R magnitudes ($-23.0 \le R \le -22.5$) and built a volume-limited sample since we want to compare the spectral properties across a uniform class of hosts.
For each class of compactness previously discussed  we evaluated the fraction $N_{em}/N_{abg}$ where $N_{em}$ is the number of emission-line systems in the class and $N_{abg}$ is the number of passive (absorption line) systems.
We assume  Poisson statistics in evaluating the errors on these fractions.

\begin{figure}[h]
\begin{center}
\includegraphics[scale=0.4, angle=90]{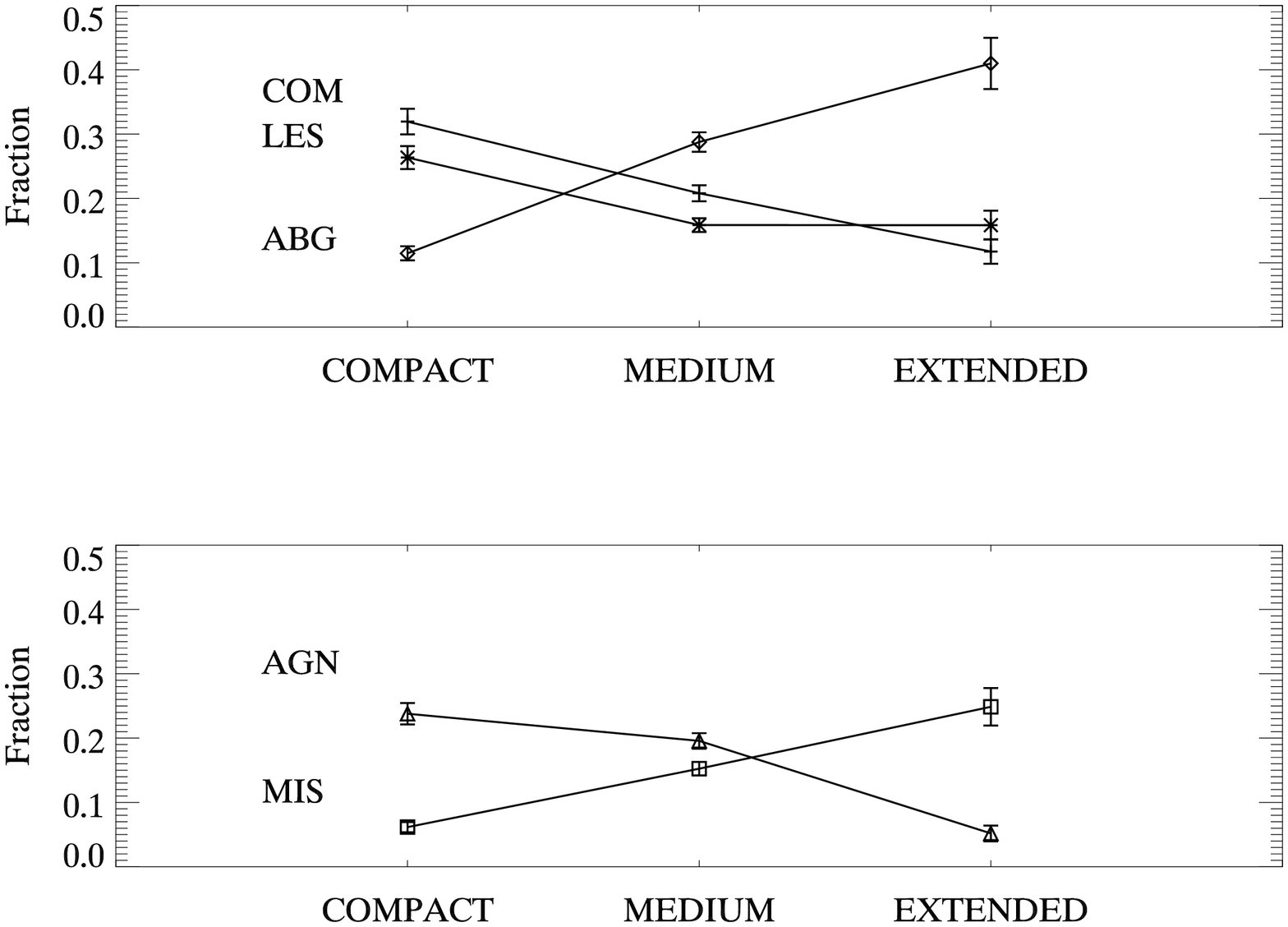}
\end{center}
\caption{\emph{Top panel: fractional abundance of composite (crosses), low-emission AGN (stars) and absorption-line system (diamonds) in different classes of radio source compactness. Error bars are calculated from Poisson statistics. Bottom panel: same as above for strong-lined AGN classified as Seyfert II or LINERs (Triangles) and low-emission misclassified systems (squares) }}
\label{compactness_spectro}
\end{figure}

For the point-like AGN we find a ratio of 4.70$\pm$0.45, for the compact jet-like AGN  1.55 $\pm$ 0.09, and for extended jets  1.25 $\pm$ 0.13.The Wilcoxon test evaluates this decline as significant  at a level of 10$^{-5}$. The fraction of emission line systems increases with the compactness of the radio AGN. The larger the jets the more unlikely are detectable emission-lines in their spectra.We discussed the lack of emission lines in galaxies powering strong jets  in Paper I and in   \cite{Reviglio2006}. Interestingly compact jets form a class between point-like and large jet sources, and  the abundance of emission line systems is more than four times higher than the abundance of absorption-line systems in hosts containing very compact radio sources. This can be seen more clearly in figure \ref{compactness_spectro} where we show the fractional abundances for different spectroscopic types  with radio AGN of different compactness. Emission-line AGN dominate the compact class and then progressively decrease. In the extended class low-signal-to-noise, weak-line AGN dominate together with spectroscopically passive galaxies.

We conclude that the radio and spectroscopic signatures of  AGN activity in radio galaxies \emph{anti-correlate}, with luminous large jets showing less remarkable lines than compact radio sources at all SMBH masses. We note that in Paper II we have shown that a weakening of the lines is correlated with a decrease in star-formation activity in the host galaxy, as well as  a change in morphology. This change is more substantial for less luminous hosts  in agreement with evidences for ``downsizing'' in both the mass assembly of early type galaxies \cite[]{Cimatti2006} and SMBH growth \cite[]{Merloni2004} and can be interpreted as the effect of a progressive exhaustion of the gas present in a galaxy in a secular evolution scenario.

If the gaseous reservoir of a galaxy is progressively transferred towards inner regions at higher rates when the galaxy is young and at lower rates when it becomes older, the accretion rate onto the SMBH would vary over time, making the AGN activity on average wane over time, in agreement with the findings of a decrease in AGN strength in X-ray and radio power from other surveys.   It would also explain the findings by \cite{Emonts2007}  that point-like radio emission is associated with hosts that have detectable neutral hydrogen, while strong jets seem to lack such gas. A progressive change in the measured accretion rates of radio galaxies has been reported by \cite{Marchesini2004}. 
A variation of the accretion over time associated with a change in the radio morphology would call for an evolutionary model of AGN activity where lines progressively decrease in strength and the radio emission transforms from point-like to jet-like.

\section{Analysis of Unified Models}
\label{Unified}

Anisotropy in the radiation of active nuclei has been well established in the past decades \cite{Urry1995} and most features of AGN are currently explained in terms of orientation of the nuclear component with respect of the observer. 

The results presented in this paper which link the size of a jet to the mass of the SMBH powering it appear problematic for such models, since the SMBH mass  is a quantity independent of orientation.

 In this section we examine the predictions of unified models using  our large statistical sample to see how well such models perform.

First we compared the masses of the black holes for radio galaxies showing  FR2 sources, selected from our narrow-line sample,   and  the masses of point-like
 quasars and  found $10^{8.37\pm 0.03} \msun$  for the
 point-like radio quasars   and $10^{8.34\pm 0.06}$ \msun for FR II, suggesting
 that these two types of objects are indeed powered by the same type of
 black holes, and confirming that quasars are mostly associated with ellipticals  \cite{Dunlop2003}  since only these systems have such massive black holes.
 This is in agreement with the unified paradigm.

We compared the radio luminosities of the two classes. If unified by
 orientation, the two classes must have the same radio luminosity, unless relativistic boosting is important. The median radio luminosity of point-like quasars in our sample is
 $10^{30.3}$ \lradio, while the median radio luminosity of FR2s associated with
 elliptical galaxies is $10^{31.6}$ \lradio. The fact that there is
 more than an order of magnitude difference in the two populations is
 inconsistent with the hypothesis that the two classes can be unified by
 orientation.  Boosting of radio emission given by orientation would increase
 the radio emission of point-like quasars, not decrease it and cannot be the cause of such difference.  The fact that these point sources are less luminous than the extended ones might suggest that their jets are intrinsically smaller.

The two populations also differ in the density of their environments,
 with a density for point-like radio quasars of 2.77 galaxies/Mpc$^3$  and
 density of the FR2 sources of 4.72 galaxies/Mpc$^3$. The environmental
 density of FR2s is almost twice as great as  the density surrounding radio
 quasars. If we restrict the analysis to point-like quasars and FR2s  with L$>10^{31}$ \lradio to avoid troubles with the sampling of the luminosity function of
 the two populations,  we find that the average density for the
 radio-loud point-like quasars is still  2.77 galaxies/Mpc$^3$  while for the FR2 is 4.35 galaxies/Mpc$^3$, in agreement with the  result by
 \cite{Prestage1983} who found  that the two population do not inhabit the same
 environments, using a completely different sample and density estimator. This is also inconsistent with the unified model since inclination should
 not depend on environmental density.

We also note that the fact that we find 13 FR2 galaxies associated with
 broad- emission-line systems is not consistent with the geometric
 interpretation: broad-line systems should not display large jets, since the
 jet should be pointing towards the observer. FR2s  are routinely found
 associated with quasars, at higher redshifts, consistent
 with our finding, a fact that  undermines the assumption that orientation
 effects can be used to explain the main differences between radio
 quasars and elliptical radio galaxies.

We also examined  the idea that Seyfert IIs  are obscured Seyfert I
 systems from a statistical point of view, since several studies have pointed
 out inconsistencies with such an interpretation, finding differences in
 the polarization properties of the Seyfert II nuclei \cite[]{Tran2003},
 differences in the host features \cite[]{Malkan1998}, and differences in the
 distribution of the [OIII] emission \cite[]{ Celotti2005}.

The underlying assumption of the unified models is that Seyfert I and
 Seyfert II systems host the same AGN, and that the two population are
 different only because of the different viewing angle with respect to an
 absorbing dusty torus.
 In order to make a fair comparison with the Seyfert I sample which was
 selected at redshifts $z<0.2$, Seyfert II galaxies used in this
 analysis  were selected in the same redshift interval, using the standard emission-line flux diagnostic $[OIII]/H\beta > 3$ for systems with signal-to-noise higher than 3.

First, we compared the X-ray properties of these systems. The fraction
 of X-ray emitting systems among Seyfert I is 30\%, while for Seyfert II
 is only 0.4\%. This is consistent with the idea that a very large
 fraction of the X-rays are obscured in Seyfert II galaxies as suggested
 by other studies. This would be in qualitative agreement with the
 expectations of the unified models, since X-ray emission would be extincted
 by the obscuring torus.

Secondly we compared the Balmer decrements for these two types of
 systems. The Balmer decrement gives an estimate of the extinction for
 systems. For AGN in the absence of obscuration the ratio should be $\sim$ 3.1 \cite[]{Osterbrock2006}.
 We find that the Balmer decrement for Seyfert Is is 3.74, fairly close
 to a non extincted system,  while for Seyfert IIs it  is 10.5, clearly
 showing that Seyfert II are dust-extincted, in agreement with orientation models.

In the orientation paradigm, a Seyfert I luminosity is given by the sum of the  luminosity of the nucleus and the host, while in Seyfert IIs most of the nuclear light is blocked by the absorbing torus.
We calculated the average reddening E(U-R) expected for these systems
 from the standard relation between the extinction $A_V$ and  median
 Blamer decrement  $A_V \sim 2.99~Log\frac{B}{B_i}$, assuming
  $B_i=3.1$. We adopted the standard slope   $R_V=\frac{A_V}{E(B-V)}=3.1$ for the
 extinction curve near the V band , and the conversion between the B and V
 bands to U and R bands from Rieke \& Lebofsky (1985). We found that the
 expected extinction  E(U-R) is ~1.2 mag. The extinction coefficients are
  respectively $A_U\sim2.4$  and $A_R\sim1.2$. If the nucleus of  a
 Seyfert I is dimmed by obscuration to produce a Seyfert II such  that
 contribution to the  optical fluxes from the host becomes predominant, one
 would expect that the luminosity of  Seyfert Is after subtraction of the
 part obscured, should be of the order of  the luminosity of a typical
 Seyfert II. 

We compared the average absolute magnitude in U and R of the Seyfert I
 sample with the absolute magnitudes of the Seyfert II sample.
 The median absolute U magnitude is $\overline{M_U}(SeyI)= -19.32$. For the
 Seyfert II sample the median absolute U magnitude is $\overline{M_U}(SeyII)>=-18.96$.
The extinction calculated is $A_U\sim 2.4$. Therefore it appears that, on average,  Seyfert Is are not bright enough to account for the extinction expected by their Balmer decrements. The situation is worse for the R-band: in this case
 we find $\overline{M_R}(SeyI)=-20.4350$ and $\overline{M_R}(SeyII)=-21.5404$. Seyfert II are brighter in R than Seyfert I, while for unified theories the
 luminosity of Seyfert Is should always  be  higher than for Seyfert IIs, since
 their light is the  sum of the host plus the light from the nucleus.
  This result is incompatible with orientation and poses a severe problem to
 orientation models.

Dust heated up by high energy photons re-radiates in the FIR. The
 assumed obscuring torus  re-radiates in all directions, and  therefore its
 orientation should not significantly affect the amount of light
 collected in a low resolution survey such as IRAS.
 If the same absorbing medium
   is present both in broad and narrow emission line systems and they
 differ only in orientation, approximately the same fraction of FIR
 emitting sources should be found in both classes of objects and the
 luminosity in the FIR should be about the same.
On the contrary, in  our sample we found a   higher fraction of
 infrared detections for the Seyfert I sample. The fraction found is
 2.6$\pm$0.5 \%, while for Seyfert II is 1.8$\pm$0.2 \%. The difference is not very large, yet statistically significant. Moreover the average luminosity of
 the FIR emitting galaxies among the Seyfert I sample is  about
 \emph{twice} as  high as that  found for Seyfert IIs at 60$\mu$m, a fact that contradicts the idea that the two classes of systems host the same type of active nucleus.

Another quantity that can be compared is the density of the environment
 for broad and narrow emission-line systems: we  find that the two
 populations inhabit similar environments, with a slight preference for
 Seyfert II to reside in denser environment: the median density for Seyfert I
 is 2.13$\pm$0.04 galaxy/Mpc$^3$ , while for Seyfert II is 2.59$\pm$
 0.02 galaxy/Mpc$^3$, in agreement with early results by \cite{Petrosian1982}.
Orientation does not depend on the environmental
 density, so this might be regarded as a potential further problem,
 although the difference in environmental density is not as strongly marked as
 for the radio quasars vs radio-galaxy sample.
We evaluated the median black hole mass for the Seyfert I population
 and obtained   $\mhole(H\beta)=10^{7.63\pm 0.03}$. For the Seyfert II
 population the values obtained is  $\mhole(\sigma)=10^{7.86 \pm 0.06}$,
 where we used the $\sigma$ estimate since its spread is significantly
 lower than the black hole R estimate. The errors bars represent the
 maximum and minimum error associated with the method estimation.  Using a
 Wilcoxon test, we find that the probability that the two  median values are  consistent with each other is   $<10^{-4}$.
Under the assumption that  the two black hole estimates are consistent
 with each other this would imply that Seyfert I host somewhat smaller
 BH masses than Seyfert II. This would be consistent with the fact they
 reside in slightly denser environments, since galaxies tend to have
 bigger bulges and therefore more massive SMBH in denser environments.

For radio emitting Seyferts we  compared the fraction of the jet to
 point-like sources. If the jets of Seyferts are oriented perpendicular to
 the obscuring torus one would expect that for the Seyfert I population
 the fraction should be smaller than for Seyfert II, since the
 former should have the jet pointing towards the observer and therefore
  more sources among them should be unresolved. On the contrary we find a
 similar ratio of jet to unresolved sources in both classes:
 1.21 for Seyfert I and 1.26 for Seyfert II.
Moreover the average jet size for Seyfert I is larger than for Seyfert
 II: the median jet size for Seyfert I is 8.66 kpc while for Seyfert II
 is 6.73 kpc.
This is also problematic for the orientation paradigm, if jets are
 mostly perpendicular to the obscuring tori.

\begin{figure}[h]
\begin{center}
\includegraphics[scale=0.5, angle=90]{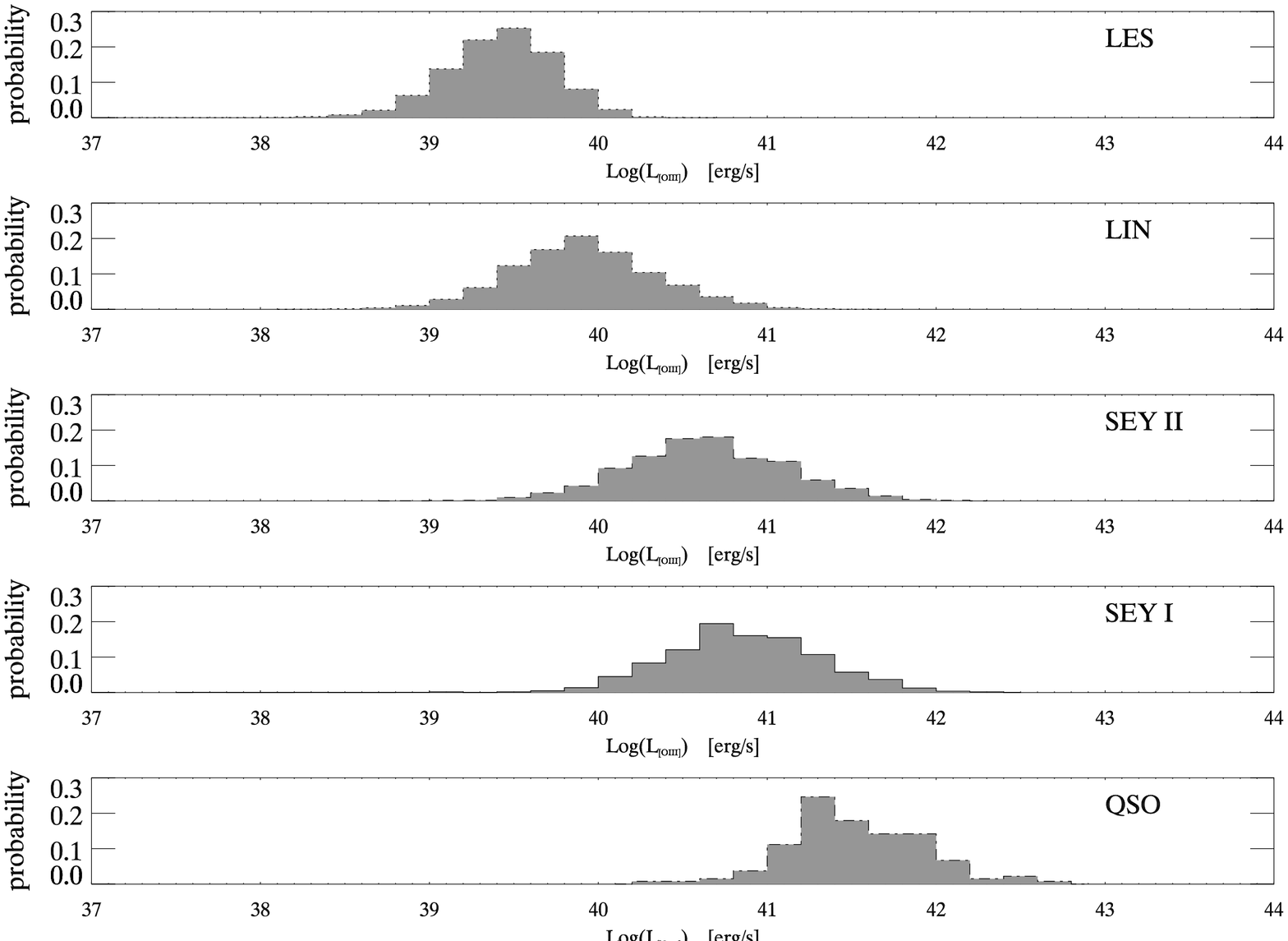}
\end{center}
\caption{\emph{Distribution of the [OIII] luminosities for different classes of spectroscopic objects }}
\label{OIII_distrib}
\end{figure}

We also confirm the difference in [OIII] luminosity distribution  for
 Seyfert I and Seyfert II sources discussed by \cite{Celotti2005}. The
 distributions  found  for different spectroscopic classes are shown in
 figure \ref{OIII_distrib}.
 The median $L_{[OIII]}$ for Seyfert I in our sample is $10^{40.8}$ erg
 s$^{-1}$, while for  Seyfert II is more than one order of magnitude
 lower: $10^{39.7}$ erg s$^{-1}$. It must be noted that  $L_{[OIII]}$
  is supposed to be an isotropic indicator not affected by the torus
 obscuration (unlike $L_{H\alpha}$) since its emission comes from the  narrow-line region which is at \emph{larger}  radii than the region where the
 obscuring torus is supposed to lie in unified models (if the torus
 should enshroud the narrow line region as well,  then   the opening angle
 for the radiation to emerge from the nucleus would be so large that the
 inclination of the system with respect to the observer would not  make
 a significant difference) . Therefore a difference in [OIII]
 luminosities between Seyfert Is and IIs further support the idea that these
 systems are physically different, with Seyfert I galaxies intrinsically more
 powerful sources than Seyfert IIs.

\begin{figure}[h]
\begin{center}
\includegraphics[scale=0.5, angle=90]{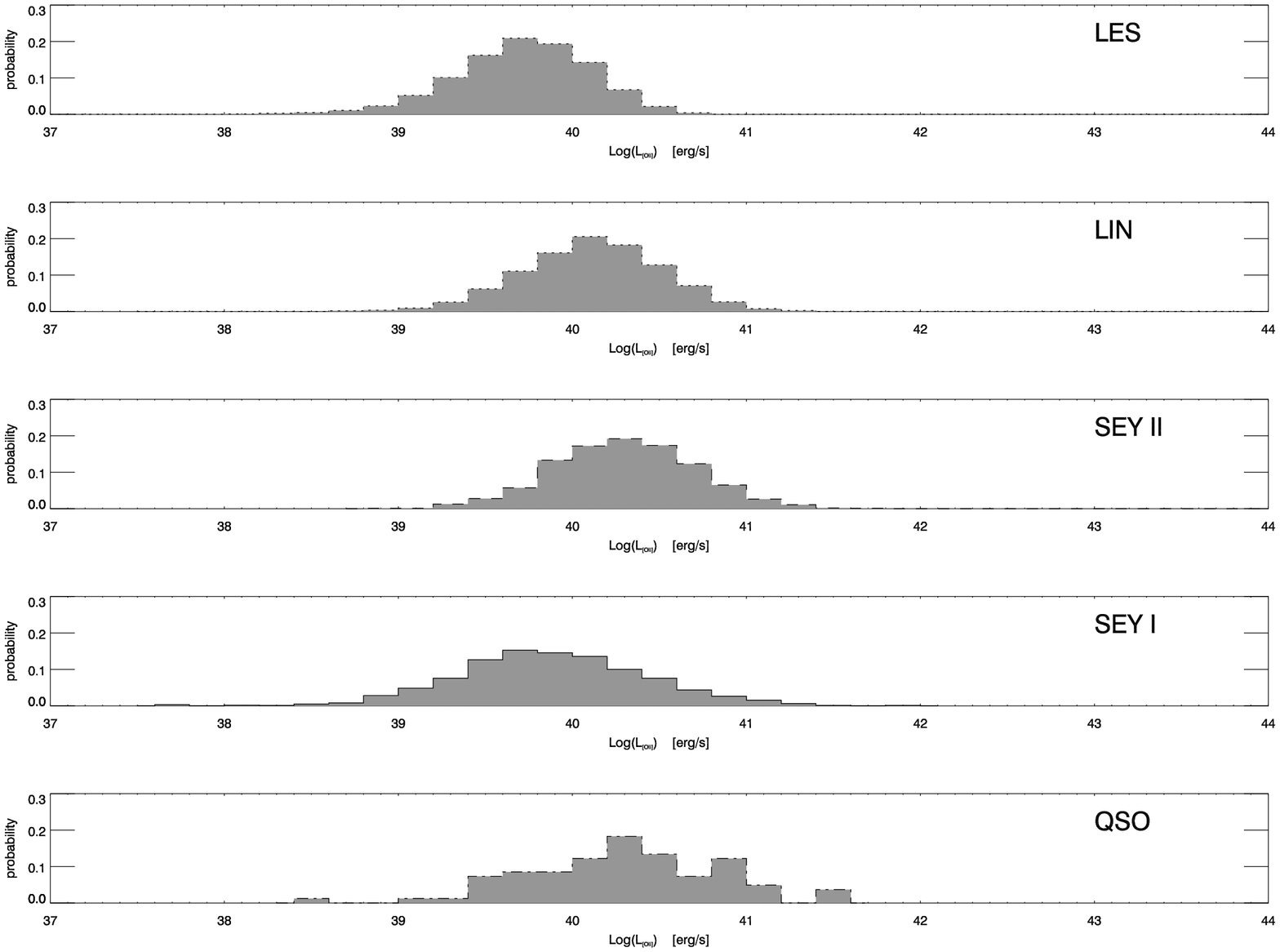}
\end{center}
\caption{\emph{Distribution of the [OII] luminosities for different classes of spectroscopic objects }}
\label{OII_distrib}
\end{figure}

This difference in the strength of  [OIII] between broad and narrow-line nuclei was noted by \cite{Jackson1990}. \cite{Hes1993}, however,  argued that [OIII] might be contaminated by nuclear star-formation and anisotropic obscuration and that [OII] would be a better indicator, claiming to show a lack in substantial difference in [OII] luminosity between the broad and the narrow line population.  In figure \ref{OII_distrib}  we show that there exists an even more marked difference in the luminosity of the [OII] between Seyfert I and Seyfert II, inconsistent with the earlier claim.  Interestingly, unlike all the other lines, we find that [OII] is stronger in Seyfert IIs than Seyfert Is. [OII] has now been shown to be strongly correlated with star-formation \cite[]{Kewley2006}, not AGN activity. It appears more likely that the difference in [OII] in Seyfert II and Seyfert I is given by  a different amount of nuclear star-formation, or by contamination by the the galaxy star-forming regions, which according to the study of \cite{Malkan1998} are more frequent in Seyfert IIs than in Seyfert Is.

\begin{figure}[h]
\begin{center}
\includegraphics[scale=0.5, angle=90]{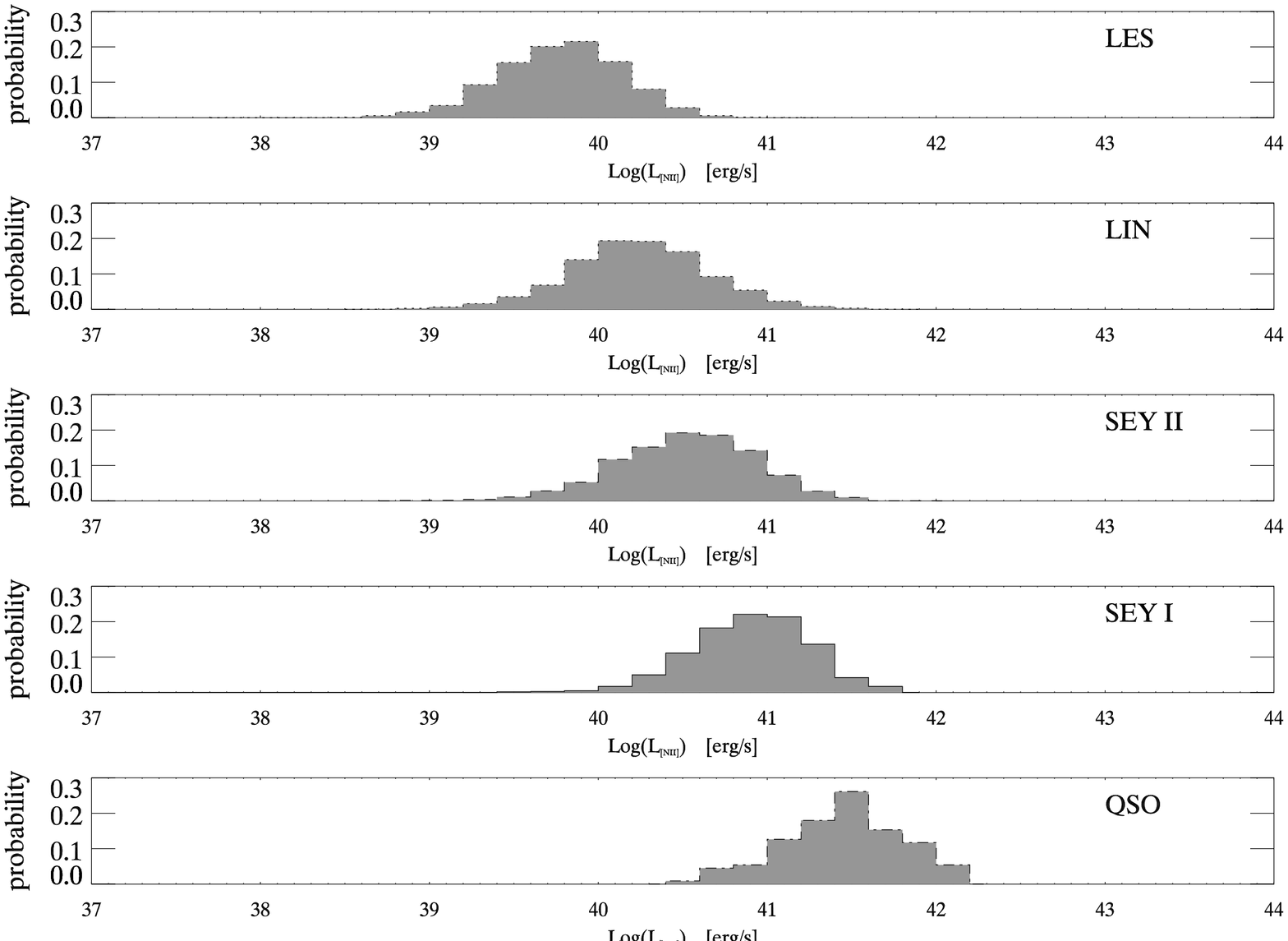}
\end{center}
\caption{\footnotesize \emph{Distribution of the [NII] luminosities for different classes of spectroscopic objects}}
\bigskip
\label{NII_distrib}
\end{figure}

\begin{figure}[h]
\begin{center}
\includegraphics[scale=0.5, angle=90]{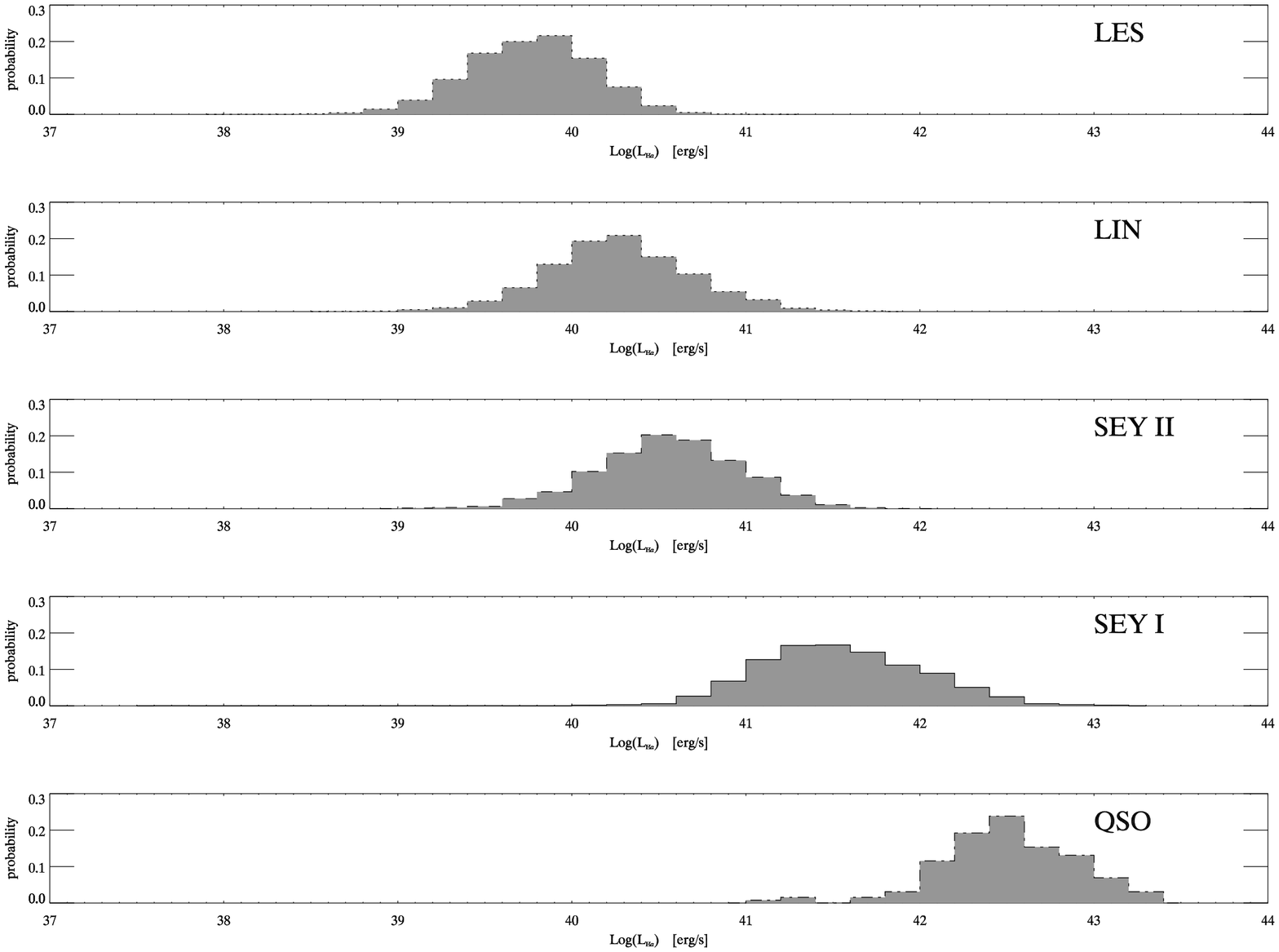}
\end{center}
\caption{\footnotesize \emph{Distribution of the H$\alpha$ luminosities for different classes of spectroscopic objects}}
\bigskip
\label{ha_distrib}
\end{figure}

Another line that can be used to confirm that different classes of AGN differ in the strength of their forbidden lines is [NII]. The distribution of the line luminosity for our sample is shown in fig. \ref{NII_distrib}. A trend similar to the one found for [OIII] is found.
Interestingly the trends found for the forbidden lines are similar to the ones found for the permitted lines: in figure \ref{ha_distrib} we show the trend for H$\alpha$ .
This suggests that strength of forbidden and permitted lines varies in a similar way across these spectroscopic classes.

In unified models, Seyfert Is and Quasars are supposed to be sources observed close to face on. When compared to narrow-lined AGN they would exhibit strong x-ray emission, point-like radio emission, bluer colors, and  less extinction.
However, how  do these two classes of putatively  face-on AGN compare to each other? The ratio of point-like sources to jet-like sources is 2.6 for Quasars and 0.8 for Seyfert I. Unresolved radio sources dominate the Quasars class, but not  the Seyfert I class. One possible effect might be given by the fact that Quasars, being hosted by more massive ellipticals are selected at higher redshift than Seyfert I and their potential jet may be more likely unresolved. If we restrict our sample to to the same redshift range  0.1$<$z$<$0.2, the results are unchanged:  we still find a stronger preference for  point-like radio sources in  Quasars than with Seyfert I. 
The other possibility, that stronger jets are powered by smaller SMBHs, is at odds with what we find in section \ref{host}.

 If Seyfert Is were AGN seen face on like Quasars, one would expect a similar fractional amount of resolved to unresolved radio sources. One might argue that  Seyfert I may be more inclined systems than Quasars. In this case one  would expect their extinction to be higher. On the contrary  the Balmer ratio for Seyfert Is is 3.8, while for Quasars it is 4.8. Nearby Quasars appear to be more extincted than Seyfert Is. 

While the data are consistent with the standard picture in which Seyfert II are dust-extincted systems while Seyfert I are not,  we find many inconsistencies with the assumption  that orientation \emph{only} can explain the differences between the two classes. Likewise, several  incongruities are found in our data  for the paradigm which suggests that nuclei of radio galaxies hosting strong jets are radio quasars seen at a larger angle.

\section{Discussion}
\label{Discussion_ch4}

In this study we have demonstrated a smooth variation in the  the radio and spectroscopic properties of broad- and narrow-line AGN  with increasing host galaxy luminosity (or alternatively baryonic mass). In particular, we showed that, with increasing host luminosity, radio morphologies become more extended  and the average radio luminosities increase; the AGN emission lines, however, become less remarkable and the extinction increases. These trends are accompanied by a variation in the properties of the host from late- to early-type, with star-formation quenching at earlier times and colors becoming redder as we  discussed in greater detail  in Paper II.

The physical properties of galaxies and their nuclei thus appear  to be well-correlated, suggesting a closer link between the properties of the nuclei of galaxies and their hosts, beyond the well-known bulge-SMBH relation \cite[]{Kormendy1992, Ferrarese2000}. 

 If late- and early-type galaxies are systems in different stages of their evolution (or transformation)  these trends may trace the evolution of active nuclei, from point-like systems with strong line emission  to jet-like systems with low or unremarkable line emission. As we move to higher redshift, the properties of AGN associated with massive SMBHs  do become more similar to the one observed in later types galaxies in the nearby universe: their lines become more prominent, they become less extincted,  their radio emission is predominantly point-like and their host become bluer \cite{Dunlop1990}. However, in unified models, these differences are mostly interpreted in terms of orientation, not evolution. In this paper however, we question the generality of such models, which do not seem to meet some simple statistical tests, as already pointed out by other authors.

The increasing evidence for downsizing in many galactic features, from star-formation to SMBH accretion requires models of AGN activity that link the properties of the  nuclei to the properties of their host galaxies.

If downsizing is simply the consequence of  a different rate in the evolution of systems of different mass (with the more massive ones evolving faster) then AGN activity would peak at different epochs for different SMBH masses as observed \cite[]{Cowie2003}, and the  SMBH  assembly would follow the  spheroid assembly  as shown  by \cite{Merloni2004}, \cite{Merloni2004b} (paper II), and \cite{Cimatti2006}. Given the well-known density-morphology correlation, this evolution would be accompanied by a change in the typical regions where different types of AGN are found at different redshifts. If the quasar phase is an early phase of nuclear activity, when a galaxy hosting a big SMBH  is richer in gas , then more massive galaxies, which inhabit denser environments, would undergo such a phase at earlier times and  quasars clustering would appear to increase with  increasing redshift, as observed \cite[]{Lafranca1998,Shen2007}.
Furthermore, as we discussed in Paper II, the star-formation rates are expected to progressively change as the host galaxy becomes depleted of cold gas, either because it is  transported into the inner regions by secular instabilities or because it is heated up by the central engine \cite[]{Ciotti2007}. If this process happens with higher speed in bigger galaxies as suggested by studies of their stellar populations, then downsizing in star-formation \cite[]{Cowie1996} would be expected to accompany the evolution in nuclear features.

Considering that high-redshift surveys are showing increasing proofs of downsizing in all galaxy features, progressive quenching of star-formation, mass assembly, AGN features and that the strength of active nuclei is well-established to evolve over time, further investigation of evolutionary scenarios of AGN activity \cite[]{Ryle1967, Lynden1969, Rees1982, Odea1998, Harvanek2001} is warranted.

Assuming that evolution is the main factor behind the photometric, morphological and spectroscopic differences in active nuclei as discussed in the previous chapters, we here discuss a speculative model of AGN activity based on spherical accretion, where  the progressive depletion of cold gas in the host galaxy leads to a progressive transformation of the nuclei. 

 This transformation would be in agreement with the evidence for co-evolution of AGN and host galaxy features presented in the previous chapters and the results from high-redshift surveys. Orientation effects can be expected in this paradigm, as we will discuss.

For semplicity, we will consider  accretion from a homogeneous gaseous reservoir that has lost most of its angular momentum. In such a case, the accretion is spherical. Recent evidence in agreement with  spherical accretion in early type galaxies has been presented by \cite{Allen2006}.

We notice that the main feature of spherical accretion is  a dependence of the accretion rates on the density of the outer regions of the gaseous reservoir from which the SMBH is accreting (Bondi 1952). In this way, it links the properties of host with the accretion properties of the nuclei, the main point we illustrated in the previous sections.

Furthermore in a spherically accreting system, a gradient in temperature and density of the gas is formed as the gas accumulates in the inner regions. The structure of an active galactic nucleus would resemble that of the interior of a protostar, with shells of different  density and temperature contracting under the influence of gravity until a stable configuration is reached.  The density profile of a spherically accreting system is typically a power law,  $\rho \propto R^{-\delta}$, with $\delta=1.5-2.0$ (Larson 1969). For SMBH of $10^9$ \msun the gas may heat up to temperatures $10^9-10^{10} $K in the inner shells, and lower temperatures at outer radii. (cf. \cite{Park1998} for a review of possible solutions). The sustainability of such gradients depends on the accretion rates and the thermodynamic properties of the gas. 

If  the radiative cooling and the heating from accretion onto a SMBH and/or a shock \cite[]{Meszaros1983} propagating in the medium  are in equilibrium, a hot temperature in the inner shells  can be sustained over time. As we will show, such a hot component, if sustained, can easily account for the high luminosity of the nuclei since it is a strong emitter of bremstrahalung radiation, independent from the fact that spherical accretion efficiencies are low.

\cite{Ostriker2005}  have pointed out that if gas surrounding 
a SMBH  intercepts 0.5\% of the radiation, its thermodynamic equilibrium would be significantly altered: if radiative heating from the accretion is effective, a hot phase of the gas surrounding the SMBH can be long-lived. Quasars are associated to deep potential wells  given their high SMBH masses and are very strong radiators: the expectation of a hot, turbulent, dense phase of the gas surrounding their SMBH seems  physically more plausible than  an ensamble of small, warm ($T\sim 10^4$ K) long-lived  clouds  nicely orbiting the SMBH  in Keplerian orbits, with low filling factor as postulated in unified scenarios. 

Radiative heating can sustain high temperatures in the surrounding regions of SMBH, well beyond the assumed $10^{-3}$ pc scale of the accretion disk. For example, if the inner regions are fed by a  cooling flow from the outer region of a galaxy   \cite[]{Quataert2000}, an increasing trend in temperature  of the gas from outer regions  (100-1000 pc) down to  radii of  $\sim$1 pc is found. At 1 pc, where their simulation stops,  the temperature reaches a  temperature    T$>10^{8.5} $ K. 

In this picture  three main regions surround a SMBH: a region filled by hot, dense gas, a region filled by  gas of medium density and medium temperatures and  a region filled by low density cool gas. The relative size of these three regions depends on the hydrodynamic  (and possibly magnetohydrodynamical) and thermodynamical  evolution of the region. As we will discuss in detail we identify these three regions as the three main components of the standard AGN picture: the broad line region (BLR), the warm absorber and the narrow line region. 

The main radiative process in a hot, dense  plasma    is thermal bremsstrahlung. The maximum total  emission from Bremsstrahlung expected in a hot plasma per unit volume in the absence of absorption  is $\epsilon=1.4\times 10^{-27}n_e n_iZ^2T^{1/2}g$ erg s$^{-1}$ cm$^{-3}$, where T is the temperature of the plasma, $n_i$ is the number density of the ions, $n_e$ is the number density of the electrons, and g is the Gaunt factor which is approximately equal to 1. For a neutral plasma $n_i \sim n_e$. For scale values typical of a BLR ($n_i\sim 10^{10}$ ions/cm$^3$, R$\sim$ 0.1 pc) and an average  temperature typical of the hot phase ($T\sim 10^9$ K), one obtains  that the bolometric emission from a region of volume $V\sim R^3 \sim (0.1 pc)^3$ may account for luminosities as high as  $L_{Brem} \sim \epsilon R^3\sim 10^{50}$ erg s$^{-1}$. This shows that this picture has the energetics to account for the high bolometric luminosities of quasars ($10^{48}$ erg s$^{-1}$), even in the presence of substantial absorption by the surrounding medium. Furthermore the flatness of the bremsstrahlung spectrum would be in agreement with the typical flatness of AGN spectrum across several bands.

The physics of dense plasmas at the high temperatures predicted by spherical accretion models is poorly understood. However, we note that for the  temperatures expected in the inner regions, one expects thermal  broadening of the recombination lines to be substantial.  Neglecting relativistic corrections, in first approximation the dispersion of the lines and the temperature are related by $\sigma\sim 0.1\sqrt{T} ~~[cgs]$. FWHM can be obtained by multiplying by $2\sqrt{2ln2}\sim 2.35$.
For temperatures of $T\sim10^9-10^{10}$ K one would obtain $\sigma \sim 3000-10000 km s^{-1}$ typical of permitted lines of quasars. 
In models of spherical accretion, smaller SMBH black holes would have lower accretion rates and therefore are expected to be able to sustain lower temperatures in their inner regions. In agreement with this one finds that for sigma of  $\sim {10^3}$ km s$^{-1}$ typical of Seyferts,  the temperature of the gas must reach $T\sim 10^8$ K, at least an order of magnitude lower than quasars.

We note that a  model where the inner regions host a hot, dense plasma may help account for the variability observed in such nuclei.

Variability in quasars is on timescales of light-years, while only  light week or days in Seyferts (Barvainis 1993). This would strengthen the point that the emitting region of quasars is on scales $\sim 0.1 pc$ while for Seyfert is at least an order of magnitude smaller, in agreement with the expectation that these nuclei can only sustain high temperatures in smaller volumes given their lower SMBH masses.

The cooling timescale for bremsstrahlung can be estimated as $\tau=5\times 10^7 n^{-1}T_8^{1/2}$ years  for n$\sim10^{12}$ cm$^{-3}$ and temperatures of  $10^9-10^{10}$K one obtains timescales of order of few  hours to days. This can simply account for the observed  variability of the emission across the whole spectrum from radio to gamma ray for quasars and blazars on timescales of hours to days: if the temperature of the gas is not smoothly sustained over timescales of hours by the central engine, the observed power of the source changes significantly.
If, on the contrary, the   temperatures fall below   $10^8 K$  or less, the broad component of emission lines would disappear, since the thermal broadening of the lines would now be on the order of narrow-line broadening; however, the bremsstrahlung continuum   would still be strong since it scales with $\sqrt{T}$. The bolometric luminosity of a hot gas sphere at T$=10^7$K with n$=10^{12}$ and r$=0.1$ would be just one order of magnitude lower than the luminosity of a T$=10^9$. Therefore one expects a hot region with such characteristics to produce a lower luminosity quasar with narrow lines. Given the lower temperatures in the gas surrounding the SMBH ,  dust survival is favored as we will discuss, and the quasar may  appear more extincted.

Halpern (1981) has suggested the existence of a warm absorber (T$\sim 10^5$) in the region in between the broad and the narrow line region of AGN which would account for the absorption features observed in the X-ray part of the spectrum. In this model the existence of a warm absorber may be simply explained, since the structure of the gas is stratified in density and temperatures.
At larger radii, the temperature and the density of the gas drops significantly and in models of spherical accretion, a  region of lower density with temperatures of $10^5-10^6$ K   surrounds  the hot region.

Regarding this region, we note that Barvanis (1993) has summarized the evidence in favor of a warm phase of the plasma, the bremsstrahlung emission of which would account for  the UV bump observed in quasars spectra. 
Reynolds et al. (1997)have discussed the presence of dust in the warm absorber, under the assumption that dust grains decouple from the plasma temperature and are kept at lower temperatures.

Dust would be in any case present in the outer regions where the temperature falls below the sublimation temperature of dust,  which is a few thousands Kelvins.  Such dust can absorb the UV radiation produced in this shell and re-emit it in the infrared, creating the infrared bump observed in spectra.
Alternatively, we have suggested that comptonization by the warm medium of low energy photons coming from the inner region may be responsible for the production of the UV bump. 
Interestingly \cite{McDowell1989} have discussed the fact that the UV bump is \emph{not} present in all quasars spectra, which suggests that it is \emph{not} a features given by accretion itself, since in such event should be present in all spectra. In the scenario that we are considering, the presence or absence of the warm phase would be responsible for the presence or absence of the UV bump. This region may be absent, and the UV bump not observed:  Krolik et al. 1981 have shown that a T$\sim 10^5-10^7$ K phase of the gas surrounding a quasars is unstable, and therefore it is likely that is short lived. Incidently this has the implication that thermal widths of order few hundreds, which would be associated to the warm phase of the gas  are not easily observed since they are short lived. In such case one would expect a lack of systems with FWHM of few hundreds, as observed in the bi-modal distribution of the permitted line widths among active nuclei, (cf. figure \ref{narrow_width_distr} and \ref{broad_width_distr}).

In this picture at radii larger than 1pc the gas may be too cold for the lines to be thermally broadened.  At $T\sim 10^4$  the $\sigma$ of the line is  of order 10 km/s, while the observed size is $10^2$ km/s.  The broadening must come from the gas random motion and possibly infall.   This region would have densities below $10^8 cm^{-3}$ and therefore permitted and forbidden lines would originate in this region. We associate this region to the Narrow Line Region of current models. The low densities allow for forbidden lines to form, while in the inner regions, the high density would not allow the associated atomic transitions. If the size of the hot gas becomes smaller and the output of high energy photons drops, then one would expect lower [OIII] in low luminosity nuclei such as Seyfer II and radio galaxies than in their high luminosity counterparts, as found in the sample analyzed in this paper and others \cite[]{Jackson1990,Celotti2005}.

Since the temperatures for dust sublimation is of order $10^3$ K, dust can survive in this region, possibly even in the warmer zone, if it decouples from the hydrogen plasma and its temperature falls below $10^3$ K.  It is on the contrary unlikely that dust would survive in the  hot phase of the plasma surrounding the central engine.   Therefore systems with broad emission-lines would show lower levels of extinction, as observed,  since dust molecules surrounding the host phase are spread over a large volume at large distances from the nucleus.  At later times, as the region shrinks and the material in the outer regions  falls in  and accumulates keeping lower temperatures, the dust density  increases and the nucleus becomes dust enshrouded. Its lines become more extincted. In association to this the broad line component disappears,  because of the lack of a substantial fraction of  hot dense  medium, and possibly because of the enhanced extinction. If the gas in this region keeps some angular momentum, the geometry of this region might be ellipsoidal and the dust distribution would be toroidal. In this way the anisotropic features observed in the radiation of AGN may be recovered and  a contribution by orientation to their features would be expected. If a fraction of the light from the inner hot shells is scattered into the line of sight, the results of spectrophotopolarimetry \cite{Antonucci1993} may be recovered. The hidden ``Type 1'' nucleus seen in polarized light, would be given by the radiation by the innermost shells scattered into the line of sight.
This way the contribution of orientation to the features of the AGN typical of unified model  can be incorporated in the model. The effect of orientation would be more significant in systems where the gas has higher angular momentum, i.e. later types galaxies.  Seyfert galaxies, which typically are hosted by later type galaxies, would show more significant orientation-driven features, as observations suggest.

In summary  the basic idea is that it is not the  efficient conversion of matter into light what produces the AGN luminosity, but the presence of relatively large volumes of hot gas in the inner 0.1 pc. If the galaxy becomes gas poor, however, the accretion rates  fall and the hot phase might not be sustained: this  would bring the hot shell to be progressively confined to smaller radii and eventually disappear, making the AGN fade and its broad emission line disappear.  At later times, since cool gas is more likely to infall than warmer gas,  the low-power AGN phase may be followed by a new cycle of  activity if the accretion is high enough to bring the surrounding gas to high  temperatures again.  This may eventually trigger another cycle of AGN activity, until the density of the gas in the outskirt is so low that the infall of material is unable to sustain a hot dense medium in the surroundings of the AGN. In such case the nucleus would still be accreting, but it would not shine. This may help explain the existence of nuclei in elliptical galaxies which  show a large amount of gas in their nuclear regions, and yet they  do not power an AGN \cite[cfr.][]{Ferrarese2005}.

A fraction of about 10\% quasars and AGN show radio emission. We note that a  stratified medium in density and temperature may explain many salient  features of the radio emission in the nuclei of galaxies.

The need for a stratified medium in the inner regions of galaxies has been suggested by Bicknell et al.  1997 to explain the features of Compact Steep Spectrum (CSS) radio sources and Gigahertz-Peaked Spectrum (GPS) radio sources. 
The main difference in these two classes is their linear size $l$ and the peak of their radio emission $\nu_m$. These two have been shown to correlate with law $\nu_m\propto l^{-0.65}$ (cfr. O'Dea 1998), with more extended sources peaking at lower frequencies.

Bicknell et al. (1997) showed that the properties of GPS and CSS sources can be described by a model involving interaction of a jet with a dense interstellar medium with density described by a power law $R^{-\alpha}$ with $\alpha\sim 1.5-2$ and density of 10--100 atoms cm$^{-3}$ at distances of 1 kpc. 
This distribution in density would give densities above $10^{10} $ atoms cm$^{3}$ in the inner 0.1 pc, in agreement with the model we discussed. The shell structure of the ISM they postulated is also  in agreement with the one we would expect  in our scenario.

 In the model discussed by Bicknell et al. (1997), the interaction of the jet and the dense interstellar medium produces a bow shock, which defines the size of the sources. As the shock front proceeds into the interstellar medium, the free-free self-absorption of the radio emission varies, being more effective when the jet traverses the denser shells, since the optical depth of the bremsstrahlung  self-absorption scales as $\propto  n_e^2 $.  
Since quasars are known to power mostly point-like sources, and yet harbor massive SMBH, at the high end of the distribution in mass of  SMBH in galaxies, this leads to the question of why these systems most of the time do not show remarkable jets, but mostly unresolved or parsec-scale jets. It also leads to the question of why these systems sometimes do power double lobed radio jets (De Vries et al.  2007).

We note that a stratified medium as the one suggested by Bicknell et al. (1997) may help explain this, if we consider that in the inner regions, high temperatures are sustained. A shell of hot, dense gas is much more effective in frustrating a jet than cold dense gas, since pressure is proportional to the temperature. 
If we consider the model adopted by Bicknell et al. (1997),  a jet is treated as an ellipsoidal cocoon with pressure $P_c$ expanding in a surrounding medium described by density profile  $\rho^{-\delta}$. The jet has a energy flux $F_E$. The work that a cocoon can do on the external medium is  $W=P_c\frac{dV}{dt}=\frac{3}{8-\delta}$. For $\delta \sim 2$ this becomes $W=0.5F_E$ and the work is proportional to the energy flux of the jet. The typical jet energy fluxes estimated by Bicknell et al. (1997) are $10^{45}-10^{46}$ erg s$^{-1}$ and are in agreement with evalutation in other studies (e.g., Celotti et al. 1993).
In the picture we are discussing, the early stages the cocoon expands in a hot and dense medium with mean temperature $T=10^9K$ (or higher)  and density $\rho=10^{12}$ cm$^{-3}$. Assuming that the jet is relativistic, the expansion timescale  for it to emerge from the hot zone is of order $t=R_{hot}/c\sim10^7 s$. The work that it needs to do to win the pressure of the external medium is of order $P_{ext}\frac{dV}{dt}\sim P_{ext}\frac{V_c}{t}$, where $V_c$ is volume of the cocoon. The external pressure from the hot plasma is $P_{ext}=nkT\sim 10^5$ dyne cm$^{-2}$. The work needed to win the external pressure would be $\frac{\pi P_{ext} R_{hot}^2}{c\xi^2} $. For $\xi=2$, i.e. a cocoon with major axis twice as long as the minor axis, and a jet that expends up to the boundaries of the hot zone, the required work is of order $2\times 10^{51}$ erg s$^{-1}$
This means that a typical jet with flux energy of order $10^{45}$ erg s$^{-1}$, would not have enough energy to break through the inner hot zone and will be confined and disrupted inside it. The radio source would appear point-like or barely resolved. If a parsec-scale jet is powered, it may be substantially decelerated by the motion through the hot gas zone: this may provide a mechanism to explain the observed jet deceleration in BLac blazar \cite[]{Wang2004}, for example.

Furthermore, only highly relativistic jets would be able to emerge from the inner regions, suggesting that, if observed, jets associated to quasars and blazars must be highly relativistic, since jets with lower kinetic energies would not be able to win the pressure of the hot region shell and would not  be launched.
This selection may explain the  high Lorentz factors of jets associated to quasars and blazars.  

A more complete discussion of this model can be found in Reviglio (2008). A detailed modeling of activity in such scenario is beyond the scope of this paper.
We note however that in the scenario described, the results differences in [OIII], [OII], [NII] shown in section \ref{Unified} , the anticorrelation between jet displacement and strength of the lines and the correlation between the properties of the host and the properties of the nuclei shown in this paper may find a simpler explanation.

\section{Conclusion}
\label{Conclusion_ch4}

We have presented a comparative study of a local population of more than 150000 narrow- and broad-lined active galactic nuclei drawn from the Sloan Digital Sky Survey (2DR) and their radio counterparts drawn from the FIRST and NVSS surveys.

After classification of radio morphology for more than 4000 sources, we have shown that more massive central black holes are more likely associated with more powerful and larger jets and that this correlation between radio power and supermassive black hole mass is found also in the fainter population of radio AGN which is not detected at the current sensitivity limit of all-sky radio surveys.

 We have compared this effect in different region of the large scale structure and found that this correlation is primarily driven by the host mass (and associated SMBH mass), not by the large scale environment in which the host galaxy reside. This supports the idea that the average size of a jet is determined by the strength of its kinetic power and not by the environment in which it propagates, in agreement with results from magneto-hydrodynamic simulations of AGN. A small, yet significant increase in the fractional amount of jets in denser environment is nonetheless found, which suggests that close interactions in denser environments  may be  capable of powering stronger radio activity.

We have compared the strength of the radio emission in our AGN  with the strength of their lines and found that the more prominent the jet, the weaker the lines. We interpret this as an effect of gas depletion. In support of this idea, we find that more compact jets are associated to galaxies with more recent episodes of star formation and therefore richer in cold gas, while large extended jets tend to be associated to systems with little ongoing star-formation and most likely gas depleted. This is in agreement with the study of Emonts et al. 2007, who find that point-like unresolved radio sources are associated to galaxies richer in HI, while FR I sources do not show detection in HI.

Assuming a simple spherical accretion scenario with gas brought into the inner regions by the secular evolution of the galaxy, the anti-correlation between the spectral signature strength and the radio signature strength may  be explained if we assume that more massive  SMBH are progressively brought to starvation by depletion of cold gas in the host: as the density of cold gas in the host drops, the SMBH would accrete at lower rates. From this, it follows that the accretion rates would be  lower, the emission of the nucleus less powerful,  the ionized regions surrounding the SMBH  thinner, and consequently the emission lines weaker. In the absence of a significant amount of dense gas,  radio jets would be less likely contained or decelerated by the interstellar medium in their inner regions  and therefore may  reach larger extents.

These findings point towards a physical link between the properties of the spectroscopic lines and the radio morphology of radio AGN, and their hosts, a fact that might question the assumption of the unified models based on orientation. Moreover, the trends in spectroscopic and radio features of nearby active nuclei appear to form a  continuum with increasing galaxy mass. This seems hard to reconcile with scenarios where most of the features of AGN are interpreted in terms of various degrees of extinction given by the orientation of the nucleus with respect to the observer.

We have tested the expectation of unified models based on orientation with the statistical  properties of  our large sample of AGN. We showed several  incongruities with the expectations of such models, which suggest that the differences of broad- and narrow-line  nuclei  cannot simply be  an effect of orientation, in agreement with studies of other authors.
In particular we find that narrow and broad-line nuclei selected in the same volume of the universe show distribution of forbidden line luminosities ([OIII], [OII],[NII]) markedly different.This is at odds  with the expectations of a simple orientation models, since the forbidden line  emission comes from regions at much larger radii than the obscuring torus assumed in such models and therefore should be similar in broad- and narrow-line AGN. A revision of such models may be necessary.

We have presented an alternative model  of AGN activity which may account for the evolution of active nuclei over cosmic times and link the properties of a host galaxy  with the properties of its nucleus.

In the model discussed, based  on a spherical accretion scenario, the features of high- and low-power AGN are explained in terms of the emission and absorption properties of  shells of  gas  surrounding a supermassive black hole.

Unlike current models of AGN, the strong emission of the AGN is not given  by the accretion itself, which in spherical accretion is known to be inefficient, but by the hot ($T>10^8$) shell of gas in the inner region of the gaseous  reservoir surrounding a SMBH, at distances $\lesssim$ 0.1 pc, coincident with the standard broad-line region of quasars and Seyfert I nuclei. If radiative heating is efficient \cite{Ostriker2005}, and the accretion rates substantial,  high temperatures can be sustained in the inner shells over timescales longer  than  $10^6$ yr. We have shown that this hot component is an efficient emitter of   bremsstrahlung radiation which, after reprocessing from the medium in which propagates, can account for many observed features of active nuclei spectra, including the broad line emission, the variability on small time scales, the UV and infrared bump. 

Outer shells would account for the other components of the standard AGN picture: the warm absorber and the narrow-line region.
In these regions, the temperature of the gas is much lower than in the innermost shells and dust can survive, and eventually be transfered into the inner regions. If the hot region is confined to smaller radii over time, dust progressively enshroud the nucleus, making low-power AGN more extincted, as observed.
If this dust component retains part of its angular momentum, its geometry may be ellipsoidal/toroidal, and geometric effects may be incorporated in the model.
Unlike unified models, where the stability of such component has been long questioned, in this model the dust component is just part of a continuum distribution of gas around the SMBH, and can be stable.

 As the galaxy becomes depleted of cold gas, the accretion rates  drop and the nuclei are brought to starvation. High temperatures may not be supported anymore and the AGN  would eventually fade over time. This transformation would be accompanied by the disappearence of the broad component of permitted lines, and a progressive increase in the dust extinction.

We noticed that a hot, dense core may help explain the radio morphologies of active nuclei, with point-like sources associated to gas rich systems and jets to gas depleted ones, since a hot core may efficiently contain and disrupt jets.

In this scenario, active nuclei would evolve from high to low power, transforming their spectral features and their radio morphologies over cosmic times, in agreement with an evolutionary path from quasars to radio galaxies envisioned by Lynden-Bell (1969)

\section{Acknowledgments}

We wish to thank Jacqueline van Gorkom, Bob Becker, Laura Kay,  Ben Johnson for helpful discussions.

PMR was supported by the National Science Foundation under grant AST-06-07643.

Funding for Sloan Digital Sky Survey project has been provided  by the Alfred P. Sloan Foundation, the Participating Institutions, the National Aeronautics and Space Administration, the National Science Foundation, the U.S. Department of Energy, the Japanese Monbukagakusho, and the Max Planck Society. The SDSS website is http://www.sdss.org .

The SDSS is managed by the Astrophysical Research Consortium (ARC) for the Participating Institutions (The University of Chicago, Fermilab, the Institute for Advanced Study, the Japan Participation Group, The Johns Hopkins University, the Korean Scientist Group, Los Alamos National Laboratory, the Max-Planck-Institute for Astronomy, the Max-Planck-Institute for Astrophysics, New Mexico State University, University of Pittsburgh, University of Portsmouth, Princeton University, the United States Naval Observatory, and the University of Washington).

\end{document}